\RequirePackage{fix-cm}
\documentclass[smallextended]{svjour3}       
\smartqed  
\usepackage{graphicx}
\usepackage{subfigure}
\usepackage{amsbsy}
\usepackage{amsmath}
\usepackage{url}

\usepackage{latexsym}

\renewcommand{\u}{\underline}
\renewcommand{\b}{\bar}

\newcommand{\bc}{\begin{center}}
\newcommand{\ec}{\end{center}}
\newcommand{\bd}{\begin{description}}
\newcommand{\ed}{\end{description}}

\newcommand{\full}{\ensuremath{\alpha d_{k,t}+b_{k} + \tau^{t-k+1}}}
\newcommand{\noPA}{\ensuremath{b_{k}+ \tau^{t-k+1}}}
\newcommand{\notau}{\ensuremath{\alpha d_{k,t}+b_{k}+1}}
\newcommand{\nobias}{\ensuremath{\alpha d_{k,t}+ \tau^{t-k+1}}}
\newcommand{\Zfull}{\ensuremath{\alpha d_{l,t}+b_{l}+\tau^{t-l+1}}}

\journalname{}
\begin{document}

\title{A likelihood-based framework for the analysis of discussion threads
}



       \author{Vicen\c{c} G\'omez         \and
               Hilbert J. Kappen          \and
               Nelly~Litvak              \and
               Andreas Kaltenbrunner 
       }
%
       \institute{Vicen\c{c} G\'omez \and Hilbert J. Kappen \at
                     Donders Institute for Brain Cognition and Behaviour,\\ 
                     Radboud University Nijmegen, 
                     The Netherlands.
                     \and Vicen\c{c} G\'omez 
                     \at\email{v.gomez@science.ru.nl}         
                     \and Hilbert J. Kappen 
                     \at\email{bertk@science.ru.nl}
                     \and Nelly~Litvak \at
                     Faculty of Electrical Engineering, Mathematics and Computer Science,
                     Department of Applied Mathematics,
                     University of Twente, The Netherlands. \\
                     \email{n.litvak@ewi.utwente.nl}
                     \and Andreas Kaltenbrunner \at
                     Information, Technology and Society Research Group, 
                     Barcelona Media,  
                     Barcelona. Spain.\\
                     \email{andreas.kaltenbrunner@barcelonamedia.org} 
       }


\maketitle

\begin{abstract}
  Online discussion threads are conversational cascades in the form of posted
  messages that can be generally found in social systems that comprise
  many-to-many interaction such as blogs, news aggregators or bulletin board
  systems. 
  We propose a framework based on generative models of growing trees
  to analyse the structure and evolution of discussion threads. We
  consider the growth of a discussion to be determined by an interplay between
  \emph{popularity}, \emph{novelty} and a trend (or \emph{bias}) to reply to
  the thread originator.  The relevance of these features is estimated using a
  full likelihood approach and allows to characterize the habits and
  communication patterns of a given platform and/or community.
%
%
%
%
%
  We apply the proposed framework on four popular websites: \emph{Slashdot},
  \emph{Barrapunto} (a Spanish version of Slashdot), \emph{Meneame} (a Spanish
  \emph{Digg}-clone) and the article discussion pages of the English \emph{Wikipedia}. 
  Our results provide significant insight into understanding how discussion
  cascades grow and have potential applications in broader contexts such as
  community management or design of communication platforms.

 

  \keywords{discussion threads \and information cascades \and
    preferential attachment \and novelty \and maximum likelihood \and
    Slashdot \and Wikipedia}
\end{abstract}

\section{Introduction}
Nowadays, online platforms where users interchange messages about a topic of
interest are ubiquitous on the Internet.  Examples range from online message
boards, blogs, newsgroups, or news aggregators to the discussion pages of the
Wikipedia.  A discussion typically starts with a broadcasted posting event that
triggers a chain reaction involving some users who actively participate in the
cascade.

\begin{figure*}[!ht]
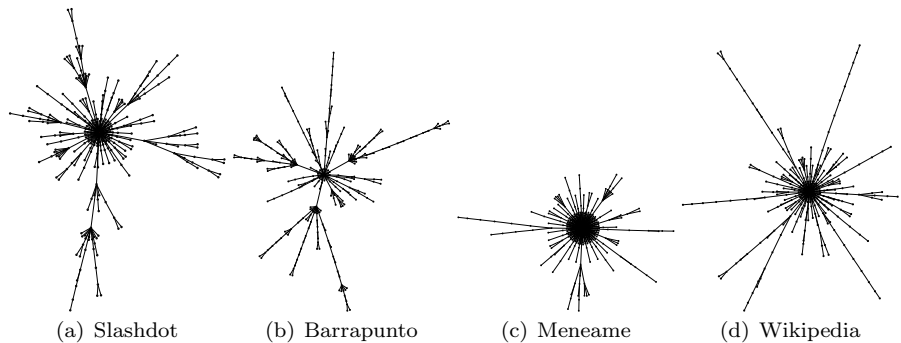

\begin{center}
 \subfigure[Slashdot]{\includegraphics[width=.24\columnwidth]{Fig1.eps}}
 \subfigure[Barrapunto]{\includegraphics[width=.24\columnwidth]{Fig2.eps}}
 \subfigure[Meneame]{\includegraphics[width=.24\columnwidth]{Fig3.eps}}
 \subfigure[Wikipedia]{\includegraphics[width=.24\columnwidth]{Fig4.eps}}
\end{center}
\caption[Example]{
Examples of discussion threads. The illustrations
represent the structure of the discussion after removing the content of the
messages. The central node corresponds to the main post (news article) and the
rest of the nodes to regular comments which are attached as replies to the
main post or to other existing comments.  Each figure corresponds to a
discussion selected randomly from each of the four websites considered in this
study.
}
\label{fig:real}
\end{figure*}

Unlike other types of information cascades, such as those corresponding to
massively circulated chain letters \cite{Nowell08}, Twitter \cite{twitter},
photo popularity on Flickr \cite{flickr} or diffusion of pages on Facebook
\cite{facebook}, where a small piece of information is just forwarded from one 
individual to another, discussion threads involve a more
elaborated interaction between users,  with uncertain (and possibly multiple)
direccions of information flow, more similar, for instance, to the cascades
extracted from phone calls \cite{peruani}.  Since threaded discussions are
in direct correspondence with the information flow in a social system,
understanding their governing mechanisms and patterns plays a fundamental role
in contexts like the spreading of technological innovations \cite{rogers},
diffusion of news and opinion \cite{gruhl,Leskovec07}, viral marketing
\cite{viral} or collective problem-solving \cite{problems}.

What determines the growth of a discussion thread?  How to predict which
comment will elicit the next reply? Is there a simple mechanism that can
capture the structure and evolution of online discussions?  To answer these
questions, it is usually believed that popular comments attract more replies,
which in turn increases their popularity, so a \emph{rich-get-richer}
phenomenon seems to play an important role.  On the other hand, given the
transitory character of certain fads, our interest decays with time, and
novelty also appears to be fundamental in determining our attention
\cite{wu2007novelty}.

In this work, we propose a modelling framework which focuses on structural
aspects of discussion threads and sheds light on the interplay between
popularity and novelty.  We introduce a parametric generative model which
combines three basic features: \emph{popularity}, \emph{novelty} and a trend
(or \emph{bias}) to reply to the thread originator.  We show that a model which
combines these three ingredients is able to capture many of the statistical
properties, as well as the thread evolution, in four popular and heterogeneous
websites.  We also use statistical tests to analyse the impact of neglecting
one of these basic features on the explanatory power of the model.  In this
way, we are able to make statistical inferences that can assess, for instance,
whether popularity is significantly more relevant than novelty in a given
web-space.

To illustrate and validate the framework, we consider four popular websites:
\emph{Slashdot}, \emph{Barrapunto} (a Spanish version of Slashdot),
\emph{Meneame} (a Spanish \emph{Digg}-clone) and the article discussion pages of the
English \emph{Wikipedia}.  These datasets are quite heterogeneous (see Figure
\ref{fig:real} for a typical thread example of each dataset).  For instance,
whereas the first three websites can be classified as news aggregators,
Wikipedia discussion pages represent a collaborative effort towards a
well-defined goal: producing a free, reliable article.  Also, at the interface
level, while Slashdot and Barrapunto provide the same hierarchically threaded
interface, Meneame provides a linear (flat) view which allows users to reply to
other users via a tagging mechanism only.  Using the same model for the four
datasets, we can segregate the heterogeneities, which are captured via the
corresponding parameter values.

\subsection{Motivation and methodology}
The aim of the present work is to propose a quantitative framework for the
statistical analysis of online discussion threads.  For that, we propose a
model that can reproduce the structural and evolving patterns of the discussion
threads of a particular website or platform.  The model considers little
semantic information.  In particular, the discussion thread is treated as a growing
network where nodes correspond to messages and links to reply actions. The
growing networks are therefore the discussions themselves, and \emph{not}
subgraphs of an underlying social network. 
Identifying a valid generative model for this type of
networks that disregards the content helps to find meaningful regularities which
uncover universal patterns and provide a fundamental understanding of
users' communication habits.

The model is a stochastic process that assumes that such patterns can be
reproduced by means of three simple features: popularity, novelty and a
trendiness to the thread initiator.  Other aspects such as the dynamics of an
underlying social network or the precise temporal timings (termination
criteria) for the discussions are not included.  These aspects could be built
"on top" of the current framework a posteriori.

Associated to each feature, there is a parameter that captures its relative
influence,
and that depends on the particular website or platform under consideration.
The framework includes a parameter estimation procedure that can be performed
independently for each dataset (a collection of discussion threads). 
Parameter estimation is based on the likelihood of the entire evolution of each
single thread of a given dataset, providing the parameterized model which
globally fits the data best.  This prevents over-fitting, that is, reproducing
very accurately particular quantities such as the number of replies per comment
(the degrees of the nodes in network terminology), at the cost of poor
approximation of other quantities.  We show that the estimation procedure is
robust, in the sense that optimal parameters are not biased, and does not require
very large datasets to be estimated.

Parameter estimates have descriptive power, since they allow the habits and
communication patterns of a given dataset to be characterised in terms of the
aforementioned features. They can also be used to establish differences between
topics or users groups within a given website.  The relevance of each of the
features is determined by the framework via model comparison, that is,
comparing the likelihood of the general model that includes all the three
features with the three reduced variants of the model that omit one of them.

\subsection{Related work}
\label{sec:related}
There is an overwhelming and vastly growing amount of literature related to
online discussion threads.  In this section, we review some of the most related
papers.
\subsubsection{Data analysis of threaded conversations}
Data from threaded conversations has been used extensively to characterize
human behaviour, for instance, to quantify how moderation affects the quality
\cite{Lampe05} and to detect social roles in Usenet \cite{fisher,Bernheim}.
Usenet is considered the first message board and the precursor of Internet
forums.  The first large-scale empirical analysis of Usenet threads was
developed in \cite{whittaker}.  The authors reported significant
heterogeneities in the levels of user participation and thread depths, and
determined meaningful correlations between certain indicators such as message
sizes, thread depths and cross-posting between groups.

Typical user behaviours have been characterised, for example, behaviours that
are dominated by responding to questions posed by other users ("answer
persons") on Usenet \cite{Welser} or lurking behaviour (the act of reading but
rarely posting on forums) on MSN bulletin board communities
\cite{Preece},\cite{Nonnecke}.  The factors that cause users who initially
posted to an online group to contribute to it again were analysed for different
platforms in \cite{Joyce2006}. In contexts of knowledge sharing, such as Yahoo
answers, interesting patterns have been found which differentiate between
discussion- and question-answer forums and their relation with the different
levels of specialisation of the users \cite{adamicyahoo}.  These studies have
important implications for cultivating valuable online communities.

From a social network perspective, the user-reply networks emerging from the
comment activity (that link two users according to their interaction) have been
analysed for bulletin board systems \cite{Zhongbao2003a,goh06}, Slashdot
\cite{gomez08}, Digg \cite{rangwala} or even Wikipedia \cite{Laniado2011}.
Although global network features  of these networks show only minor
discrepancies to other social networks, e.g. friendship networks, a rigorous
comparative analysis revealed fundamental differences in the practice of
establishing reply and friendship links in the case of Meneame, a Digg-like
website \cite{kaltenbrunner2011NRHM}.

At the thread-level, visualisation techniques of the conversations have
facilitated the understanding of the social and semantic structures
\cite{Sack,smith02}. Statistical analysis of the threads has made it possible
to identify the distinctive properties of online political conversations in
relation to other types of discussions \cite{GonzalezBailonJIT2010}, or to
derive measures that can improve the assessment of information diffusion
\cite{Mcglohon2}, popularity prediction
\cite{kaltenbrunner_LAWEB2007,Lerman,szabo} or controversy \cite{gomez08}.  
\subsubsection{Information cascades}
Recently, the term \emph{information cascades} has been adopted to describe
similar phenomena.  It has been introduced in the economic sciences
for the analysis of herding behaviour, when an individual adopts/rejects a
behaviour based on the decisions of other individuals \cite{Abhijit,Sushil}.
In the case of discussion threads, a user adopts a behaviour by actively
participating in the conversation.

The increasing availability of electronic communication data has prompted
extensive empirical work on information cascades.  The diffusion patterns seem
to depend on the nature of the cascades under consideration.  For instance,
while a Twitter study suggested that cascades spread very fast and are
predominantly shallow and wide \cite{twitter}, photo popularity on Flickr seems
to spread slowly and not widely \cite{flickr}.  Another study found that fan
pages on Facebook are triggered typically by a substantial number of users and
are not the result of single chain-reaction events \cite{facebook}.

Empirical analysis of email threads has been the subject of intense analysis
and controversy.  The diffusion patterns of two large-scale Internet
chain-letters were analysed in \cite{Nowell08}. The authors concluded that,
rather than fanning out widely and reaching many people in a few steps,
chain-letters propagate in a narrow and very deep tree-like pattern.  This
result seems to contradict the way a small-world network would operate, and
several hypotheses have been proposed to account for this observation while
preserving the small-world intuition.  One of them is the \emph{selection bias}
hypothesis, which states that the observed structures may not be typical
instances of the processes that generated them, but instead exceptional
realizations \cite{golub}.  Recently, another study \cite{Wang2011} reported
that cascades composed of forwarded emails fan out widely and quickly die out.
Despite the differences of the different studies, however, certain regularities
are pervasive in all datasets, for instance, that the largest cascades occur
with very small probability and affect a very small proportion of the whole
population.

Theoretical model analysis to understand these phenomena usually considers the
underlying networks where cascades originated. In \cite{wattscascades} two
cascading regimes which show rare but very large cascades are identified,
depending on the network connectivity: for sufficiently sparse connectivity,
cascade sizes follow a power-law distribution at a critical point, while for
sufficiently dense connectivity, cascade sizes follows a bimodal distribution.
The analysis also concludes that endogenous heterogeneities in the underlying
network (high threshold or degree variability) has mixed effects on the
likelihood of observing global cascades.

Attempts to find the underlying connectivity of associated networks using
epidemic models have been made using data from blogs \cite{gruhl,adar05}. The
conversation cascades of blogs have been considered in \cite{Leskovec07}, with
special emphasis on the scale-free character of related distributions such as
cascade sizes or degree distributions. A simple, parameter-free model able to
generate power-law distributed cascade sizes and temporal patterns resembling
the real-world ones was proposed in \cite{blogs}.  However, the role the
underlying social network plays in information diffusion also seems to be
dependent on the particular domain.  Whereas a study about social influence
concluded that diffusion of content strongly depends on the network topology
\cite{Bakshy}, email forwarding seems to be less dependent \cite{Wang2011}.
Despite the existing discrepancy about the role of network topology, it is
believed that network topology strongly determines diffusion at a microscopic
level, in the beginning of the cascade only.  At a macroscopic scale, after a
critical propagation threshold is reached, network topology does not seem to be
much relevant.

If one disregards the underlying social network which generates the cascades,
the simplest phenomenological model for cascades is a branching process, where
a random number of descendants is generated at each time step (or generation),
for each node, according to a fixed probability distribution which is equal for
all nodes in the cascade.  Galton-Watson processes are a particular type of
branching processes, and have been suggested in \cite{golub} to support the
selection bias hypothesis, in \cite{Sadikov} to account for missing information
in the cascades and in \cite{Wang2011} as baseline models.  However,
branching processes are insufficient for our purposes, mainly because they
assume each node (comment) to be independent, and therefore do not provide a
basis for the evolution of the cascade.  Instead, we are interested in the
stochastic process governing the cascade growth.

\subsubsection{Discussion threads as information cascades}
As stated previously, the aforementioned cascades involve the forwarding of a
piece of information from one individual to another one.  Discussion threads
involve a more elaborated interaction between users, with uncertain (and
possibly multiple) directions of information flow.  More recently, \cite{Kumar}
proposed a model for conversation threads which combines popularity and
novelty. The model improves on the simple branching process and qualitatively
reproduces certain statistical properties of the resulting threads and
authorships, and is illustrated using data from three popular forums, with
special emphasis on Usenet.  Similarly, \cite{gomezHT} showed that a growth
model based on a modified preferential attachment which differentiates between
the root of the thread and the rest of the nodes captures many statistical
quantities associated with the structures and the evolution of the empirical
threads.  We build on these previous works and compare extensions of both
models, providing a unifying likelihood-based framework for their parameter
estimation and validation.  Our approach allows the interplay of the different
parameters to be analysed and is validated in detail for four datasets.

\subsection{Outline}
In the next section we introduce our framework and present growth
models of discussion threads and their parameterisation. Section
\ref{sec:lik} describes our likelihood approach for parameter 
estimation and its validation.  In Section \ref{sec:results}, we
present the empirical results of this work.  First, we provide a
global description of the threading activity in the four datasets
under study in subsections \ref{sec:datasets} and \ref{sec:global}.
Our analysis also
highlights the importance of repetitive user participation in relation
to other types of cascades and their impact on the entire social
network.  We compare the explanatory power of the different proposed
models in subsection \ref{sec:comparison}. Validation of the structure
and evolution of the model generated threads is analysed in detail in
\ref{sec:structure} and \ref{sec:evolution} respectively.  Finally, in
Section~\ref{sec:discussion} we discuss the results and implications
of this work. In the Appendix we provide an analytical deviation of
the limit behaviour of the proposed models and details of
the parameter estimation procedure.
\section{Growing tree models for discussion threads}
\label{sec:model}
Before we introduce the formal model, we provide first the required mathematical terminology. 
We consider an abstract representation of a discussion thread as a graph, where
nodes correspond to comments and links between nodes denote reply actions.  The
initial (root) node has a special role: it corresponds to the triggering event of the discussion (a news
article, for instance) ad we will refert to it in what follows often as the ``post''.  
We model the growth of such a graph, in which new
nodes are added sequentially at discrete time-steps.  We consider the case that
comments are single-parental, that is, the same comment cannot be a reply of
more than one comment.  In this way, the resulting graph is a tree (it does not
contain cycles).  
The total number of nodes, or size of the discussion, is denoted by $N$.

\begin{figure}[!t]
\begin{center}
 \includegraphics[width=.5\columnwidth]{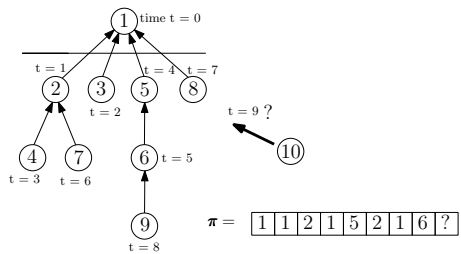}
\end{center}
\caption[Example] {Small example of a discussion thread represented
as a tree: at time-step $t=9$, node (comment) number $10$ is added to the
thread.  At the bottom right we show the corresponding vector of parents
$\boldsymbol{\pi}$. Each node attracts the new comment with different
probability according to the model under consideration (see text).  }
\label{fig:model}
\end{figure}

A compact way to represent trees consists of a vector of parent nodes that we
denote by $\boldsymbol{\pi}$.  We use the indices of the vector
$\boldsymbol{\pi}$ as the identifiers of the comments and elements of
$\boldsymbol{\pi}$ correspond to identifiers of the replied comments.
In this way, $\pi_t$ denotes the parent of the node with identifier
$t+1$, which was added at time-step $t$.  See Figure \ref{fig:model} for an
illustration.

The growth of the tree is characterized by the probability that existing nodes
attract new ones. Thus we are interested in the probability of node $k$ being
the parent $\pi_{t}$ of node $t+1$ given the past history, that we denote as
the vector $\boldsymbol{\pi}_{(1:t-1)}$. Such probability can be written as
$p(\pi_t = k| \boldsymbol{\pi}_{(1:t-1)})$, for $t>1$, $k \in \{1,\hdots,t\}$.
The vector $\boldsymbol{\pi}$ at time-step $t=1$ contains only the first reply
to the root and is denoted as $\boldsymbol{\pi}_{(1)}=(1)$
\footnote{ At time $0$ we have $\boldsymbol{\pi}_0=()$ and for all
  trees, $p(\pi_1=1) = 1$ and $0$ otherwise, i.e.
  $\boldsymbol{\pi}_1=(1)$ always.}. Note that by construction, $\pi_t
\leq t, \forall t$.

We define a growing tree model by means of its associated \emph{attractiveness}
function $\phi(k)$ (to be defined later) for each of the nodes.  Generally:
\begin{align}
\label{eq:prob}
p(\pi_{t} = k| \boldsymbol{\pi}_{(1:t-1)})
&= 
\frac{\phi(k)}{Z_t}, &
Z_t & = \sum_{l=1}^{t}{ \phi(l)},
\end{align}
where for clarity we have omitted the dependency of $\phi(k)$ and $Z_t$  on
the thread history $\boldsymbol{\pi}_{(1:t-1)}$. The term $Z_t$ is just a
normalisation sum which ensures that at every time-step, the probability of
receiving a reply is normalised and adds up to one.


Once we have introduced the stochastic process governing the thread evolution,
we present the three features that determine the attractiveness of a node.

\vspace{.1cm}
\noindent\textbf{Popularity} :
Comments receive new replies depending on how much replies they
already have.  This mechanism, known as preferential attachment (PA)
or as \emph{Mathew effect} in social sciences, has a long tradition to
characterize many types of complex networks.  Its origins date back to
the early twentieth century~\cite{polya}. More recently, PA became
popularised in the model of Barab\'asi~\cite{barabasi99a} to explain the
scale-free nature of degree distribution in complex networks.

At time $t$, we relate the \emph{popularity} of a comment with its number of
occurrences in the vector of parents.  Mathematically, the degree of a node $k$
is its number of links (degree $d_{k,t}$) before node $t+1$ is added:
\begin{align}
d_{k,t} ( \boldsymbol{\pi}_{(1:t-1)})&=
\begin{cases}
1+\sum_{m=2}^{t-1}{\delta_{k\pi_m}} & \text{for $k\in\{1,\hdots,t\}$}\\
0 & \text{otherwise},
\end{cases}
\label{eq:degree}
\end{align}
where $\delta$ is the Kronecker delta function.  In the following, we omit the
explicit dependence on $\boldsymbol{\pi}_{(1:t-1)}$, so that $d_{k,t}\equiv
d_{k,t}( \boldsymbol{\pi}_{(1:t-1)})$. 
Note that we consider an undirected graph, and every existing 
node has degree equal to one initially.

To parametrize the popularity, we introduce a weight $\alpha$ common to
\emph{all} the degrees. This factor captures the relevance of the popularity
during the growth of the tree, so a value of $\alpha$ very close to zero would
mean almost no influence of popularity and its relevance will be proportional
to $\alpha$. This model corresponds to a \emph{linear} PA model.

\vspace{.1cm}
\noindent\textbf{Novelty}:
Either because of saturation or competition, old comments gradually become less
attractive than new ones. 
We model the novelty of comment $k$ as an exponentially
decaying term:
\begin{align}
n_{k,t} &= \tau^{t-k+1},& \tau&\in[0,1].
\label{eq:novelty}
\end{align}
Note that empirical evidence exists that novelty in online spaces decays slower
than exponentially \cite{wu2007novelty,iribarren} and is strongly coupled with
circadian rhythms \cite{poisson}.  Decay in novelty also depends on the data,
e.g. news fade away rapidly compared with video popularity \cite{szabo}.
However, since we use comment arrivals as time units, these heterogeneities are
alleviated and an exponential decay is justified. In \cite{Kumar} the same
mechanism is also proposed.

\vspace{.1cm}
\noindent\textbf{Root bias}:
Finally, we explicitly distinguish between the root node of a thread and the
regular comments. On many platforms users are more inclined to start a new
sub-thread than to reply to an other comment. A convenient way to establish
such a difference is to assume that the root node has an initial popularity,
parameterised with $\beta$, which acts as a bias.  The bias of a node $k$ is
either zero or $\beta$:
\begin{align}
b_{k}&=
\begin{cases}
    \beta   & \text{for $k=1$}\\
    0       & \text{otherwise}.
\end{cases}
\label{eq:bias}
\end{align}

\subsubsection{Attractiveness function}
We define the attractiveness $\phi(k)$ of a comment $k$ as the sum of the
previous parameterised features. The interplay or relative importance
between them is determined by the concrete values (to be estimated given the
data) of the different parameters: $\alpha, \tau$ or $\beta$.
 
We propose a model that combines all the features, and name it full model~(FM).
For comparison, we also consider three reduced variants which miss one of them.
We denote the model without popularity as NO-$\alpha$, the model without
novelty as NO-$\tau$ and the model without bias as NO-bias.  According to our
formulation, the three reduced models are nested within the full model.  Table
\ref{tb:models} shows the four models we consider and their respective
attractiveness function $\phi(k)$ together with the corresponding parameter set
and constraint.  Note that the NO-$\tau$  only requires the knowledge of node
degrees and the distinction between the root and the rest of the nodes.
Including the novelty term $\tau$ makes the process dependent on the full past
history.
\begin{table}[!ht] \centering
\caption{The four models considered in this work.}{
\begin{tabular}{|c|c|c|c|}
\hline
	Model                                               &
	Attractiveness function $\phi(k)$                &
	Parameters $\boldsymbol{\theta}$										&
	Constraint \\[3pt]\hline\hline
Full model (FM)                            &    $\full$  
        & $\{\alpha,\tau,\beta\}$ & \\[3pt]\hline
Model without popularity (NO-$\alpha$)     &    $\noPA$
        & $\{\tau,\beta\}$        & $\alpha=0$  \\[3pt]\hline
Model without novelty (NO-$\tau$)          &    $\notau$
        & $\{\alpha,\beta\}$      & $\tau=1$ \\[3pt]\hline
Model without bias (NO-bias)               &    $\nobias$ 
        & $\{\alpha,\tau\}$       & $\beta=0$ \\[3pt]\hline
\end{tabular}}
\label{tb:models}
\end{table}


Other variants of this model are possible: in \cite{gomezHT}
popularity is modelled as a sub-linear PA process where the parameter $\alpha$
is exponentiating the degree and no novelty term exists.  We found no
significant differences between the FM model introduced here and a
more general model with an extra parameter exponentiating the degrees.
Thus the conclusions derived here are general and do not depend on whether a
linear o sub-linear PA process is used to model popularity.  The proposed
formulation is more convenient mathematically, since the normalisation constant
$Z_t$ does not depend on the particular structure of the thread.  For the FM,
we have:
\begin{align}
\label{eq:z}
Z_t &= \sum_{l=1}^{t}{ \Zfull} = 2\alpha(t-1) +\beta+\frac{\tau(\tau^t-1)}{\tau-1}.
\end{align}
This allows to derive the asymptotic properties of certain quantities of
interest, such as degree distributions.  If one neglects the bias to the root
node and considers instead a termination parameter $\gamma$ independent of the
thread structure, one recovers the T-MODEL proposed in \cite{Kumar}, which is
also based on a linear PA. The NO-bias model can thus be used to illustrate the
T-MODEL in the datasets considered here.


\section{Likelihood-based approach for parameter estimation}
\label{sec:lik}
We explain here our approach to find parameter estimates given a set of data.
In the following, a dataset denotes a generic collection of threads. It can
include the entire set of conversations extracted from a particular website
such as Wikipedia, but also conversations focused on particular topic domain,
for instance, the domain \emph{science} in Slashdot.

Typically, existing approaches for parameter estimation of evolving graph
models require certain assumptions to be hold.  For instance, the parameters
of a PA process in large networks are usually measured by calculating the rate
at which groups of nodes with identical connectivity form new links during a
small time interval $\Delta t$ \cite{jeong,blasio}.  However, this approach is
suitable only for networks with many nodes that are stationary in the sense
that the number of nodes remain constant during the interval $\Delta t$. This
is not a reasonable assumption in our data, which is often produced by a
transient, highly non-stationary response.

Another approach for parameter estimation relies on fitting a measured
property, for instance the degree distribution, for which an
analytical form can be derived in the model under consideration.  For
the PA model, extensive results exist with emphasis precisely on the
degree distributions \cite{Rudas,bennaim}. Following the standard
heuristic (see e.g.~\cite[Chapter 8]{Hofstad}), we obtain the
following power law behaviour of the degree distribution in FM:
\begin{equation}
\label{eq:dd}
c_1 x^{-2}\le P(\mbox{degree}\ge x)\le c_2 x^{-2},\quad 0<c_1<c_2,
\end{equation}
where $c_2$ depends strongly on $\tau$. The derivation of this result is
provided in Appendix~\ref{app:asympt}. We see that the power law exponent of the cumulative
distribution function equals 2 and does not depend on the model parameters.
Furthermore, we see from the derivation that the difference between $c_1$ and
$c_2$ can be several orders of magnitude. Thus, our results, obtained by
existing analytical methods, are too rough to enable a statistical evaluation
of $\tau$. Finally, the parameter $\beta$ does not affect (\ref{eq:dd}), but we
will see from the experiments that this parameter defines the shape of the
distribution for lower values of the degrees. We note that the analytical
derivation for the NO-$\tau$ model leads to the power law exponent
$2+1/\alpha$, which does depend on $\alpha$ but this dependence is not
prominent enough to accurately evaluate $\alpha$ from the power law exponent
estimation on the data.


Our approach considers instead the likelihood function corresponding
to the \emph{entire} generative process (instead of particular
measures such as degree distributions or subtree sizes) introduced
before. We can assign to each observation (each node arrival in each
thread) a given probability using equation \eqref{eq:prob}.  The
parameters for which the probability of the observed data is maximised
are the ones that best explain the data given the model assumptions
(see \cite{lik} for a similar approach for other network growth
model).

Formally, we observe a set $\Pi:=\{\boldsymbol{\pi}_1, \hdots
\boldsymbol{\pi}_N\}$ of $N$ trees with respective sizes
$|\boldsymbol{\pi}_i|$, $i\in\{1,\hdots N\}$ and we want to obtain estimates
$\boldsymbol{\hat\theta}$ which best explain the data $\Pi$.  If we assume that
the threads in the dataset are independent and identically distributed, the
likelihood function can be written as:
\begin{align}
\label{eq:lik}
\mathcal{L}(\boldsymbol{\Pi}| \boldsymbol\theta) 
&=\prod_{i=1}^{N}{p(\boldsymbol{\pi}_i| \boldsymbol\theta) }\notag\\
&=\prod_{i=1}^{N}\prod_{t=2}^{|\boldsymbol{\pi}_i|}{p(\pi_{t,i}|\boldsymbol{\pi}_{(1:t-1),i},\boldsymbol\theta) }\notag\\
&=\prod_{i=1}^{N}\prod_{t=2}^{|\boldsymbol{\pi}_i|}{\frac{\phi(\pi_{t,i})}{Z_{t,i}}}\quad,
\end{align}
where $\boldsymbol{\pi}_{(1:t-1),i}$ is the vector of parents in the tree $i$
after time $t-1$ and $Z_{t,i}$ is the normalisation constant $Z_t$ for thread
$i$.  We can apply this approach to each of the model variants presented in the
previous section by choosing the attractiveness function $\phi$ accordingly.  Numerically, instead
of maximising \eqref{eq:lik} directly, it is more convenient to use the
log-likelihood function.  We consider the following error function to be
minimised:
\begin{align}
\label{eq:loglik}
-\log \mathcal{L}(\boldsymbol{\Pi}| \boldsymbol\theta) & =
-\sum_{i=1}^{N} \sum_{t=2}^{|\boldsymbol{\pi}_i|} 
\phi(\pi_{t,i}) - \log {Z_{t,i}}.
\end{align}
\subsection{Validation of the Maximum Likelihood estimation procedure}
\label{sec:validation}
\begin{figure}[!t]
\begin{center}
\includegraphics[width=.7\columnwidth]{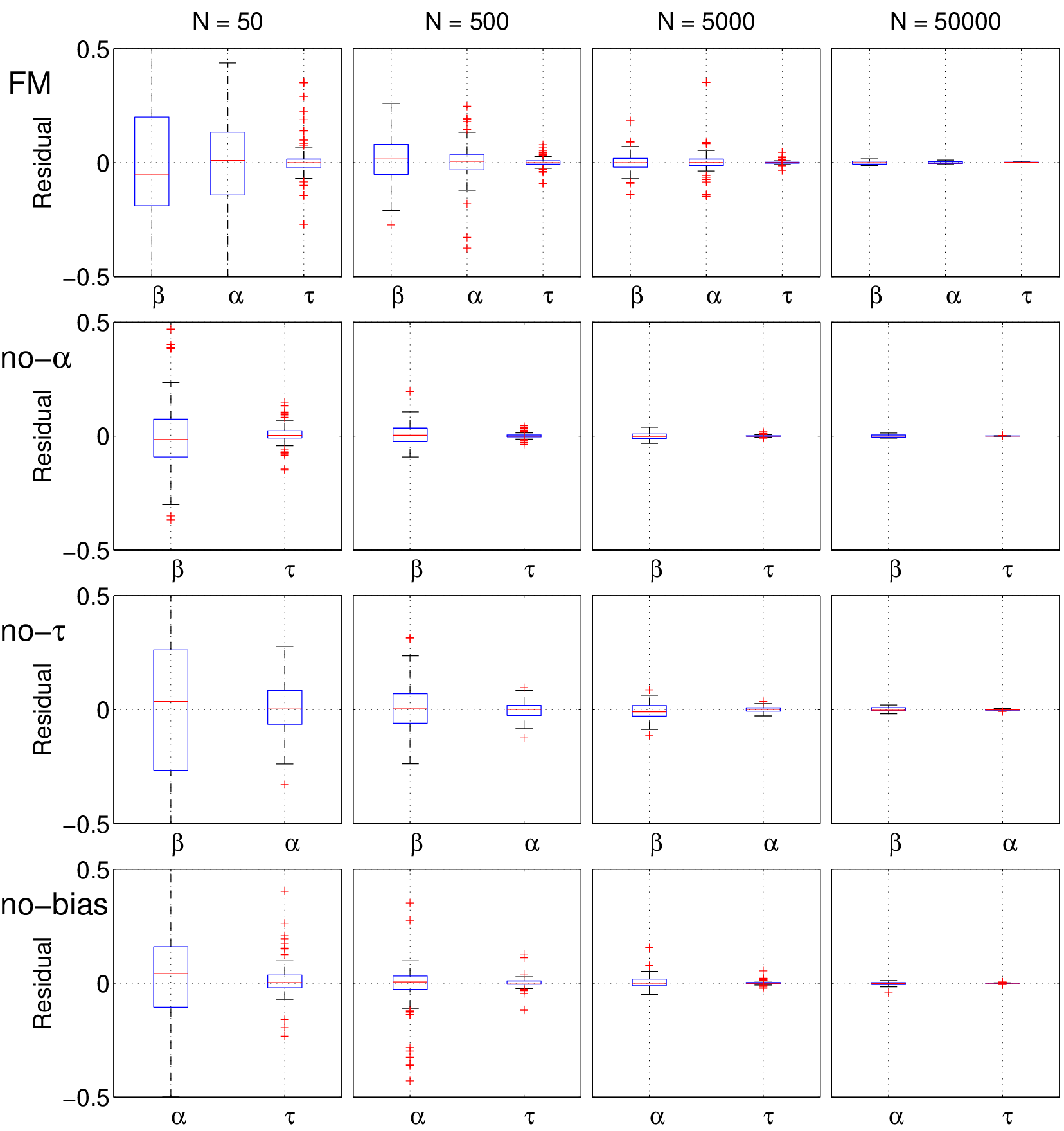}
\end{center}
\caption[Example]{
Validation of the maximum likelihood estimates: box plots of the residuals
(differences between estimates $\boldsymbol{\hat{\theta}}$ and real
values $\boldsymbol{\theta}^*$) for synthetic data.  Columns indicate 
number of threads $N$ and each row corresponds to one model (see table
\ref{tb:models}).  Data represents the outcome of $100$ independent experiments
with $\boldsymbol{\theta}^*$ selected randomly.  Estimates were initialised
using five different random initial conditions to test for multiple local
minima. The selected solution was the one corresponding to the best likelihood. 
}
\label{fig:valid}
\end{figure}

In this section we show numerically that the parameter estimates found using
the previously described optimisation are correct, i.e. not biased.  We proceed
as follows: for a given model, we generate $N$ synthetic threads with randomly
chosen parameter values $\boldsymbol{\theta}^*$ and calculate the estimated
parameters $\boldsymbol{\hat{\theta}}$ on the synthetic set via maximisation of
\eqref{eq:loglik}.  If the residuals (defined as
$\boldsymbol{\theta}^*-\boldsymbol{\hat{\theta}}$) go to zero as a function of
$N$, our estimates are non-biased.

We also tested for different local minima using five different random
initialisations. In practice, we only experienced different optimal solutions
(local minima) for small $N$, and every time this happened, each solution had a
different likelihood, which shows evidence that the optimisation problem is
well defined.


Figure \ref{fig:valid} shows box plots of the residuals. Columns indicate
number of samples $N$ and each row corresponds to one of the models under
consideration.  We can see that all models asymptotically converge to the true
values, since the residuals are practically zero for large enough $N$.
Overall, outliers (red crosses) are most frequent in the FM
estimates, which is the model with most number of parameters.  In contrast, the
model without novelty term, NO-$\tau$, is the one which shows the most stable
behaviour.  This occurs because for the other models, estimating $\tau$  is
difficult for small values ($\tau^*<0.5$) and small $N$, since novelty decays
exponentially. We will see later that for our four datasets this is not a
problem.  Interestingly, although for $N=50$ the residuals are broadly
distributed, their medians are centred at zero, which implies that even for
small number of threads, one can get a fair estimate of the optimal values.

We therefore can conclude that the proposed maximum likelihood method is
unbiased and that it is possible to obtain good parameter estimates using a few
hundreds of threads.
\section{Empirical results}
\label{sec:results}
In this section we first describe the datasets we consider and then give a
brief overview about some general characteristics.  The datasets which we
consider contain complete information of the thread evolution and are therefore
not prone to selection bias.  A summary of the datasets statistics can be found
in Table~\ref{tb:stats}.
\subsection{Description of the datasets}
\label{sec:datasets}

\begin{itemize}
\item \textbf{Slashdot (SL)} ({\url{http://slashdot.org/}}): Slashdot is a
popular technology-news website created in 1997 that publishes frequently short
news posts and allows its readers to comment on them.  Slashdot has a community
based moderation system that awards a score to every comment and upholds the
quality of discussions \cite{Lampe05}.  The interface displays hierarchically
the conversations, so users have direct access to the thread structure. A single
news post triggers typically about 200 comments (most of them in a few hours)
during the approximately 2 weeks it is open for discussion. Our dataset
contains the entire amount of discussions generated at Slashdot during a year
(from August 2005 to August 2006).  See~\cite{gomez08} for more details about
this dataset.

\item \textbf{Barrapunto (BP)} ({\url{http://barrapunto.com/}}): Barrapunto
is a Spanish version of
Slashdot created in 1999. It runs the same open source software as Slashdot,
making the visual and functional appearance of the two sites very similar. 
Although Slashdot currently runs a more sophisticated interface than Barrapunto,
they both shared the same interface at the time the data were retrieved.
They differ in the language (audience) they use and the content of the news
stories displayed, which normally does not overlap. The volume of activity on
Barrapunto is significantly lower.  A news story on Barrapunto triggers on
average around 50 comments. Our dataset contains the activity on Barrapunto
during three years (from January 2005 to December 2008).

\item \textbf{Meneame (MN)} (\url{http://www.meneame.net/}) Meneame is the most successful
Spanish news aggregator. The website is based on the idea of promoting
user-submitted links to news (stories) according to user votes. It was
launched in December of 2005 as a Spanish equivalent to \verb|Digg|.
The entry page of Meneame consists of a sequence of stories recently
promoted to the front page, as well as a link to pages containing the
most popular, and newly submitted stories. Registered users can, among
other things: \emph{(a)} publish links to relevant news which are
retained in a queue until they collect a sufficient number of votes to
be promoted to the front page of Meneame,
\emph{(b)} comment on links sent by other users (or themselves),
\emph{(c)} vote ({\it menear}) comments and links published by other users.
Contrary to both BP and SL, Meneame lacks an interface for nested comments,
which are displayed as a list. 
However, the tag \verb|#n| can be used to
indicate a reply to the $n$-th comment in the comment list and
to extract the tree structures we analyse in this study.  To focus on the most
representative cascades, we filter out stories that were not promoted, that is
marked as discarded, abuse, etc.  Our dataset contains the promoted stories
and corresponding comments during the interval between 
Dec. 2005 and July 2009.

\item \textbf{Wikipedia (WK)} ({\url{http://en.wikipedia.org}}) : The English
Wikipedia is the largest
language version of Wikipedia. Every article in Wikipedia has its
corresponding \textit{article talk page} where users can discuss on
improving the article.  For our analysis we used a dump of the English
Wikipedia of March 2010 which contained data of about 3.2 million
articles, out of which about 870,000 articles had a corresponding
discussion page with at least one comment. In total these article
discussion pages contained about 9.4 million comments. Note that the 
comments are never deleted, so this number reflects the totality of
comments ever made about the articles in the dump. The oldest comments
date back to as early as 2001. Comments who are considered a reply to
a previous comment are indented, which allows to extract the tree
structure of the discussions. Note that Wikipedia discussion pages 
contain, in addition to comments, structural elements such as
subpages, headlines, etc. which help to organize large discussions. We
eliminate all this elements and just concentrate our analysis on the
remaining pure discussion trees. More details about the dataset and
the corresponding data preparation can be found in~\cite{Laniado2010}. 
For our experiments we
selected a random subset of $50,000$ articles from the entire
dataset. Results did not vary significantly when using different
random subsets of the data.
\end{itemize}



\begin{table}[!t] \centering
\caption{Dataset statistics.}{
\begin{tabular}{|c|r@{,}l|r@{,}l|r@{,}l|c|}
\hline 
    dataset &
    \multicolumn{2}{|c|}{\#threads} &
    \multicolumn{2}{|c|}{\#nodes} &
    \multicolumn{2}{|c|}{total users} 
\\\hline
SL    &   $9$ & $820$         &   $2,028$ & $518$    &  $93$ & $638$     \\\hline
BP    &  $7$ & $485$          &  $397$ & $148$       &  $6$  & $864$    \\\hline
MN    &  $58$ & $613$         &  $2,220$ & $714$     &  $53$ & $877$    \\\hline
WK    &  $871$ & $485$        &  $9,421$ & $976$     &  $350$ & $958$   \\\hline
\end{tabular}}
\label{tb:stats}
\end{table}

\subsection{Global analysis}
\label{sec:global}
\begin{figure}[!t]
\begin{center}
\includegraphics[angle=-90,width=.8\columnwidth]{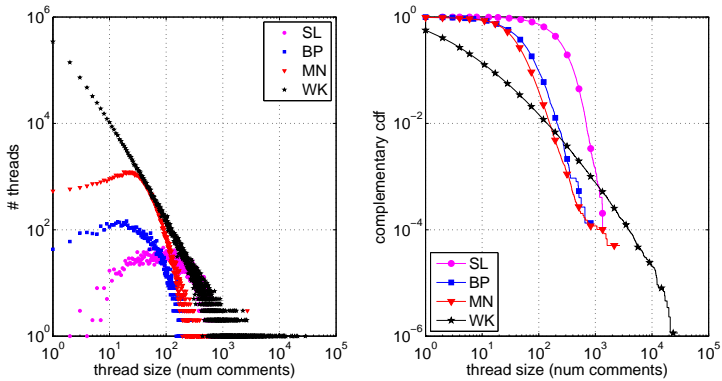}
\end{center}
\caption{
Thread sizes for the different datasets.  (left) Histogram of the sizes.
(right) Complementary cumulative distribution of the sizes.}
\label{fig:sizes}
\end{figure}
To globally characterize the threads, we analyse some properties related to the
sizes of the threads (number of comments they receive) and the authorships.

Figure \ref{fig:sizes} shows histograms of the thread sizes (left)
and their complementary cumulative distributions (right).
As expected, all distributions are positively skewed, showing a high
concentration of relatively short threads and a long tail with large threads.
However, although all distributions are heavy tailed, we clearly see a
different pattern between the three news aggregators and the Wikipedia.
Whereas SL, BP and MN present a distribution with a defined scale, the
distribution of thread sizes of Wikipedia is closer to a scale-free
distribution, in line with the threads found in weblogs \cite{Leskovec07} and
Usenet~\cite{Kumar}.  We remark that, even in the Wikipedia case, the power-law
hypothesis for the tail of this distribution is not plausible via rigorous test
analysis: we obtain an exponent of $2.17$ at the cost of discarding $97\%$ of
the data.

We also observe a progressive deviation from websites with a well defined scale
such as Slashdot, which could be described using a log-normal probability
distribution, toward websites with less defined scale such as Meneame, which
may show a power-law behaviour for thread sizes $>50$.  Barrapunto falls in the
middle and, interestingly, is more similar to Meneame than to Slashdot.

The previous considerations imply that, in general, a new post in Slashdot can
hardly stay unnoticed and will propagate almost surely over several users.
Conversely, most of the news in Meneame will only provoke a small reaction and
reach, if they do, a small group of users. We can say that, according to the
behaviour of the thread sizes, Meneame is the news aggregator that shares
most similarities with Wikipedia.
\begin{figure}[!t]
\begin{center}
\includegraphics[angle=-90,width=.8\columnwidth]{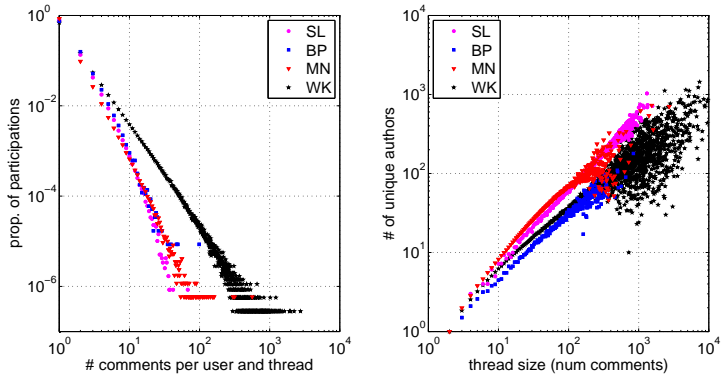}
\end{center}
\caption{Authorship: (Left) distribution of the number of comments
per user and thread in the different datasets. (Right) relation between sizes
and average number of distinct authors per thread.}
\label{fig:ComPerThread}
\end{figure}


A characteristic feature of discussion threads, unlike other form of
information cascades, is the repeated user participation.  To end this section,
we briefly mention some properties related to the authors of the comments.
Figure~\ref{fig:ComPerThread} (left) shows the distribution of the number of
participations per user and thread for the four datasets.  Although the
proportion of participations with only one comment in the discussions is large,
a significant number of participations involve at least two
or more comments. The proportion of these participations lies between 15\% for
Meneame and 31\% in the case of Wikipedia. Occasionally, some users participate
ten, hundreds or even thousands of times in the same discussion thread.

It is also interesting to analyse the relation between the size of the threads
and the authorships. This is depicted in figure Figure~\ref{fig:ComPerThread}
(right).  Although we observe a close linear relationship in all datasets as
reported for Usenet in \cite{Kumar}, we can differentiate a small decay in the
gradient present in MN and WK only, indicating that the proportion of users
that comment at least twice in the same thread becomes larger in larger
threads, something that does not seem to happen for SL at all.  We observe that
the frequency of participation on the WK talk pages is significantly higher
than the rest.

Figure \ref{fig:real} illustrates the different types of threads which we
found. We plot representative threads with similar sizes selected randomly from
each of the four datasets. For Slashdot we can see that the chain reaction is
located mainly on the initiator event (direct reactions), but some nodes also
have high degree, resulting in bursty disseminations. We could say that after a
news article is posted, the collective attention is constantly drifting from
the main post to some new comments which become more popular.  In Barrapunto we
observe similar structures, although their persistence is less noticeable.  On
the contrary, Meneame is characterised by having high concentration of nodes at
the first level together with rare but long chains of thin threads.  This
represents a pattern where only a few comments receive multiple replies, but
that sporadically can trigger a long dialog between a few users.  We note that
this phenomenon might be caused by the fact that the thread tree and, more
importantly, the number of replies a comment receives are hidden in the
interface of Meneame.  Finally, the case of Wikipedia is very similar to
Meneame, but with even longer, more frequent and finer threads of nodes with
very low degree.

\subsection{Comparison of models and interplay of the features}
\label{sec:comparison}
In this section we compare the explanatory power of the different proposed
models using the datasets previously described.  These results allow to
characterize the interplay between novelty, popularity and root bias for a
particular website. In order to compare models, we perform two types of
statistical tests based on the likelihoods.
\begin{figure}[!t]
\begin{center}
\includegraphics[angle=-90,width=.9\columnwidth]{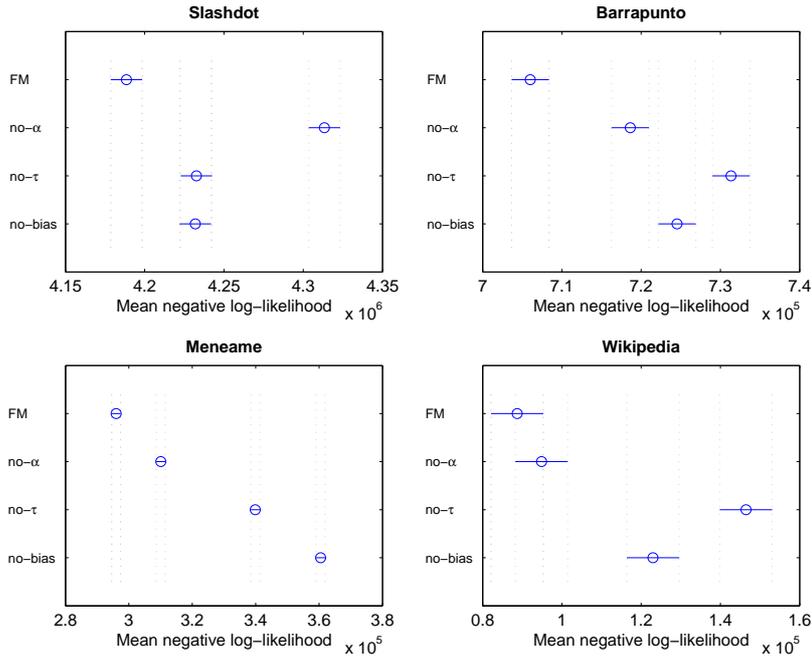}
\end{center}
\caption[Bootstrap likelihoods]{
Model comparison for each dataset:  horizontal axis shows negative
log-likelihood (left is better) and vertical axis indicates the four different
models we consider.  Two models are not statistically different if their range
plots overlap, for instance, models NO-$\tau$ and NO-bias in
Slashdot.  Conversely, model A is preferable than model B if their range plots
do not overlap and A is positioned on the left of B.  The full model
FM outperforms any of the reduced models except NO-$\alpha$
in Wikipedia.  The best reduced model depends on the dataset.  Range plots are
computed via one-sided ANOVA and Tukey's range test on the mean of Likelihoods
across $100$ (bootstrap) samples. The number of sampled threads is
$5\cdot10^4$ for all datasets.
}
\label{fig:bootstrp_lik}
\end{figure}

We first check whether the full model is significantly better than any of the
reduced models by means of a likelihood ratio test.
Results show that for all datasets except for the
Wikipedia, the full model is preferable over any of the three reduced models.
For the Wikipedia discussions, the full model, although is better than
NO-$\tau$ and NO-bias, it is not significantly different when
compared against the model without popularity (NO-$\alpha$).  This
result is important, since it highlights the main difference between the three
news websites and the Wikipedia discussion pages, namely, whereas popularity
plays a role in the news aggregators, it does not in the Wikipedia. 

To compare the reduced models, we can say that the model with better likelihood
is preferable only if the hypothesis that the likelihoods are significantly
different holds. To test whether the likelihood between the different groups
(models) differ significantly, we perform a one-sided ANOVA test and
subsequently, a Tukey's range test for multiple comparisons.  The results are
shown on Figure \ref{fig:bootstrp_lik} for each dataset.

For Slashdot, we observe that the models NO-bias and
NO-$\tau$ are not significantly different.  More importantly, the
model NO-$\alpha$ is the one which performs significantly worst.  This
indicates that neglecting the preferential attachment mechanism has the
strongest impact.  We can therefore conclude that popularity is the most
relevant feature of Slashdot: users tend to write to popular comments more than
to novel ones, for instance.

Interestingly, the PA mechanism seems to be crucial only for Slashdot. Although
one would expect very similar characterisation for Barrapunto, the relevance of
novelty and popularity differs. For Barrapunto, we observe that all the four models are
statistically different.  In decreasing order of accuracy, we have FM,
NO-$\alpha$, NO-bias and finally NO-$\tau$.  The impact of removing the novelty
term $\tau$ is therefore larger than removing the root-bias, and both features
are more relevant than the popularity.  We can conclude that the novelty is the
most relevant of the three features in Barrapunto. In contrast to Slashdot
users, Barrapunto users tend to write preferably based on how new a comment is
than how popular a comment is.

The case of Meneame is also different. The results show that the key feature to
describe Meneame is the difference between the process of writing to the post
and the process of writing to regular comments.  After this distinction is
made, we can also say that novelty is more relevant than popularity, thus in
Meneame, as in Barrapunto, users write preferably to new comments than to
popular ones.

Finally, for the Wikipedia discussion pages, as noted before, we see that
models FM and NO-$\alpha$ are not differentiable (in accordance with the
likelihood ratio test) and second, that novelty is more relevant than
differentiating between the article and the comments.

\begin{table}[!t] \centering
  \caption{Average parameter estimates for the full model over 100
      different samples and two different sample sizes. Values within
      parenthesis indicate the standard deviation of the estimated
      parameter.}
   {\begin{tabular}{|c|c|c|c|c|}
 \hline
  Dataset   &   $\log\beta$    &   $\alpha$   &  $\tau$\\\hline
  \multicolumn{4}{|c|}{$N=50$}\\\hline
  SL        &   $2.39\quad (0.17)$  &   $0.31\quad(0.02)$     &
  $0.98 \quad(0.02)$\\\hline
  BP        &   $0.93\quad (0.12)$  &   $0.08\quad(0.04)$     &
  $0.92 \quad(0.00)$\\\hline
  MN        &   $1.66\quad (0.16)$  &   $0.03\quad(0.01)$     &
  $0.72 \quad(0.04)$\\\hline
  WK        &   $-0.21\quad (0.81)$  &   $0.00\quad(0.00)$    &
  $0.40 \quad(0.19)$\\\hline
  \multicolumn{4}{|c|}{$N=5000$}\\\hline
  SL        &   $\mathbf{2.39}\quad (0.01)$  &
  $\mathbf{0.31}\quad(0.01)$     &  $\mathbf{0.98} \quad(0.00)$\\\hline
  BP        &   $\mathbf{0.96}\quad (0.02)$  &
  $\mathbf{0.08}\quad(0.00)$     &  $\mathbf{0.92} \quad(0.00)$\\\hline
  MN        &   $\mathbf{1.69}\quad (0.03)$  &
  $\mathbf{0.02}\quad(0.00)$     &  $\mathbf{0.74} \quad(0.01)$\\\hline
  WK        &   $\mathbf{0.39}\quad (0.22)$  &
  $\mathbf{0.00}\quad(0.00)$     &  $\mathbf{0.60} \quad(0.01)$\\\hline
  \end{tabular}}
\label{tb:params}
\end{table}

In the following we will contrast these conclusions with an analysis
of the parameters of the full model. In Table~\ref{tb:params} we can
find their values for two different sample sizes. We observe that
for a small sample size of $N=50$ threads, we already obtain a reliable
estimation. Only the bias to the root term ($\beta$) for Wikipedia
shows larger fluctuations. Using larger subsets of $5000$ threads does
not change the mean parameter estimates significantly, except again
for the case of $\beta$ in Wikipedia.  Thus we can conclude that the
estimated parameter values are stable using different, sufficiently
large random subsets of the data.\footnote{Note that this also
  indicates that a cross-validation (train-test) procedure would yield
  to very similar parameter estimates in train and test set.}

If we compare the actual parameter values among the different datasets
we observe results that confirm the previous conclusions like an only
minor influence of the age of a comment (novelty $\tau$ close to 1,
indicating a very slow decay) but a large impact of popularity
($\alpha$) in SL and, on the contrary, a zero value for $\alpha$ and
the biggest dependency on novelty in WK. Furthermore, if we look at
the parameter $\beta$ we find that it is largest for SL and smaller
for BP and WK. This bias to the root parameter is most important at
the beginning of a discussion where it determines whether new comments
go mainly to the root node or are replies to already existing
comments. We would expect thus to have initially broader trees for SL
(and to a minor degree for MN) while BP and WK should experience a
faster initial growth. We will analyse this point further in section
in Section~\ref{sec:evolution}.

\subsection{Model validation: structure of the threads}
\label{sec:structure}
\begin{figure}[!t]
\begin{center}
\includegraphics[angle=-90,width=\columnwidth]{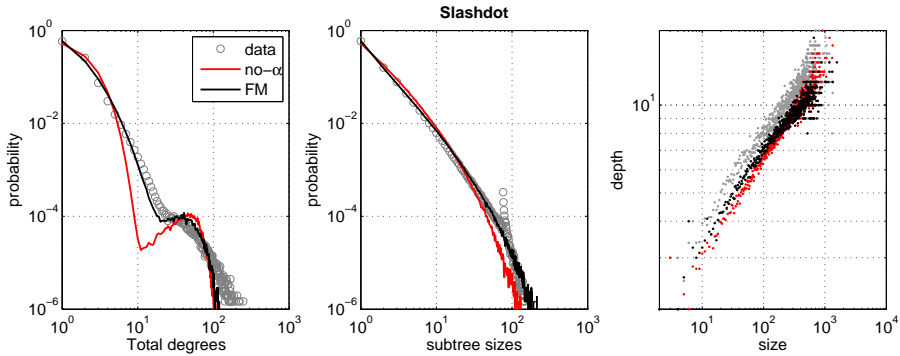}
\end{center}
\caption{Comparison of the degree distribution (left), subtree sizes
  (centre) and correlation between depth and number of comments
  (right) for the original discussion trees (gray circles) from the
  Slashdot dataset and synthetic trees generated with the full model
  (black curves and dots) or a model without popularity term (red
  curves and dots).}
\label{fig:sl_fit}
\end{figure}
To compare the real and the synthetic threads, we focus on the following
structural properties, which are calculated for both types of threads:
\begin{itemize} 
    \item \emph{Degree probability distribution:}
    we consider the probability distribution the degrees, which is equivalent
    to distribution of the number of direct replies minus one.  For this
    calculation we use all nodes, including root and non-root nodes.

    \item \emph{Subtree sizes distribution:}
    for each non-root node, we compute the probability distribution of the
    total number of its descendants, i.e. the size of the conversation triggered
    by a comment. We discard the root node because its associated distribution
    is precisely the one we use to generate the thread sizes.

    \item \emph{Relation between the sizes and depths:}
    we analyse the thread depths as a function of the thread size by taking the
    average depth of all threads with a certain size.
\end{itemize}

\begin{figure}[!t]
\begin{center}
\includegraphics[angle=-90,width=\columnwidth]{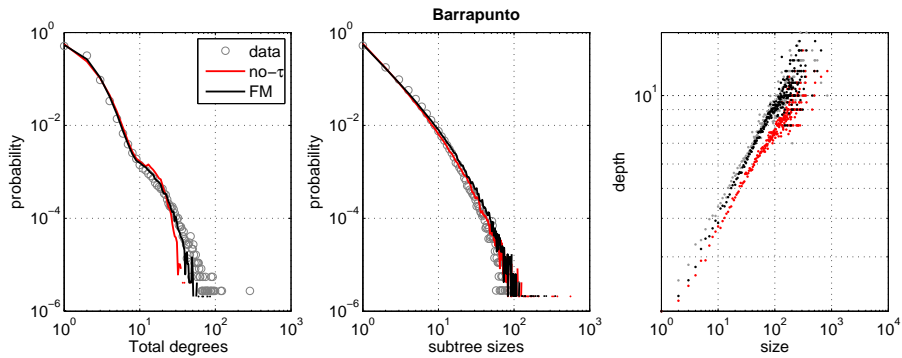}
\end{center}
\caption{Comparison of the degree distribution (left), subtree sizes
  (centre) and correlation between depth and number of comments
  (right) for the original discussion trees (gray circles) from the
  Barrapunto dataset and synthetic trees generated with the full model
  (black curves and dots) or a model not using the novelty term (red
  curves and dots).}
\label{fig:bp_fit}
\end{figure}

\begin{figure}[!t]
\begin{center}
\includegraphics[angle=-90,width=\columnwidth]{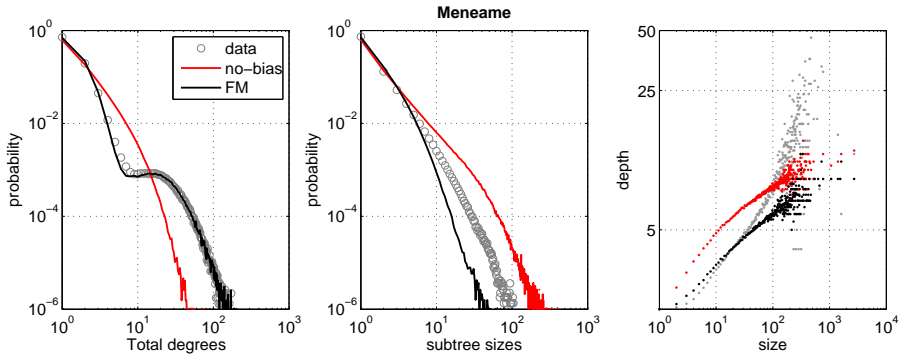}
\end{center}
\caption{Comparison of the degree distribution (left), subtree sizes
  (centre) and correlation between depth and number of comments
  (right) for original discussion trees (gray circles) from the
  Meneame dataset and synthetic trees generated with the full model
  (black curves and dots) or a model not using the bias term (red
  curves and dots).}
\label{fig:mn_fit}
\end{figure}

\begin{figure}[!t] \begin{center}
\includegraphics[angle=-90,width=\columnwidth]{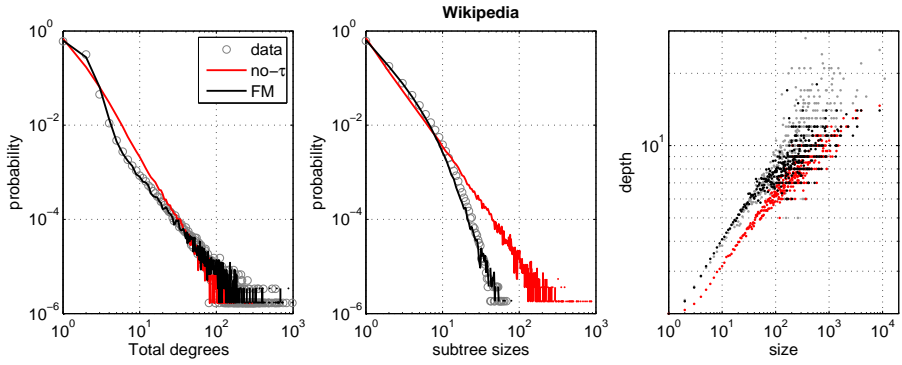}
\end{center}
\caption{Comparison of the degree distribution (left), subtree sizes
  (centre) and correlation between depth and number of comments
  (right) for original discussion trees (gray circles) from the
  Wikipedia dataset and synthetic trees generated with the full model
  (black curves and dots) or a model not using the novelty term (red
  curves and dots).}
\label{fig:wk_fit}
\end{figure}

Figures \ref{fig:sl_fit}, \ref{fig:bp_fit}, \ref{fig:mn_fit} and
\ref{fig:wk_fit} show the three previous properties for each dataset
independently in log-log axis.  For clarity, they are illustrated for the best
(FM, black lines) and the worst (red lines) model only.

Overall, the full model is able to capture reasonably well the relevant
quantities in all datasets.  In particular, the degree distributions are very
accurately reproduced, even though each dataset exhibits a different profile
(see left plot of the figures).  The effect of using a bias term is clearly
manifested in the bi-modality of that distribution, with a first peak dominated
by the comments' replies and a second peak dominated by the direct replies to
the root.  This effect is strongest in Meneame (figure \ref{fig:mn_fit} left)
and less pronounced in Barrapunto or Wikipedia, in agreement with the analysis
of the previous section, since the bias term is fundamental in the former and
less relevant in the latter datasets.

The effect of neglecting the root-bias term is that the weights of popularity
and novelty are increased and decreased respectively with respect to the
weights obtained for the FM. This effect strengthens the PA process and results
in degree and subtree sizes distributions that are too skewed for the non-root
nodes.
This issue seems to be the main limitation when the global (direct reactions to
the root) and the localised replies are not differentiated, such as in the
T-MODEL \cite{Kumar} and is closely related with the so-called stage dependency
found in \cite{Wang2011} and modelled using a branching process.

The full model also generates correct subtree sizes of the non-root nodes in
all datasets, with the exception of Meneame, which we postulate is caused by
the particularities of the platform. 
With the exception of NO-bias, all models reproduce adequately this quantity.
For the Wikipedia, however, we also observe that the model without novelty
NO-$\tau$ also produces longer tails (figure \ref{fig:wk_fit}, middle).  Since,
as we have shown before, Wikipedia can be characterised using bias and novelty
only irrespective of popularity, neglecting any of these two crucial features
affects dramatically the approximation quality of the model.

The third quantity we compare is the average depth as a function of the size of
the threads. We can see that the full model reproduces very accurately this
quantity for Wikipedia and Barrapunto, but only qualitative agreement is
reported for Slashdot. For Meneame, it clearly differs from the
data, especially for large threads.

Capturing the thread depths correctly is difficult, as pointed out in
\cite{gomezHT}, where it was shown that a model without novelty was unable to
reproduce accurately the mean depth distribution in any of the datasets we
consider here.  Indeed, figures \ref{fig:bp_fit} and \ref{fig:wk_fit} (red
curves) show that the model NO-$\tau$, which is comparable to the one
of \cite{gomezHT}, clearly underestimates the depths.  We see that the novelty
term incorporated in the full model substantially the depths to be very close
to the empirical observations in all datasets.

It is interesting to compare the observed distributions with the ones reported
in other studies. The discussion threads analysed in this work are very similar
to the conversations of Usenet \cite{Kumar}, although the relevance of the bias
term seems to be smaller for Usenet than for any of our datasets. Compared to
chain-letters, discussion threads are much shallower than the chain-letters
analysed in \cite{Nowell08} (median depths are of around $500$ levels), but
much deeper than the trees extracted from forwarded email in \cite{Wang2011}
(max depth found was four).  We have to keep in mind that the type of
interaction considered in \cite{Wang2011} and \cite{Nowell08}
differs substantially from ours.

One could consider whether the power-law hypothesis is a valid explanation for
the relation between thread sizes and thread depths as suggested for Usenet
\cite{Kumar}, or for the degree distributions as advocated for blogs data
\cite{Leskovec07}, for instance. In our datasets, we observe that rigorous
statistical tests systematically rejects the power-law hypothesis for either
degrees or subtree sizes, or does not reject it at the cost of discarding
almost all of the data.  Further, average depth does not cover more than two
orders of magnitude and are very noisy in the tail (very few posts exist with
large depths), complicating a phenomenological explanation using power laws.

\begin{figure*}[!t]
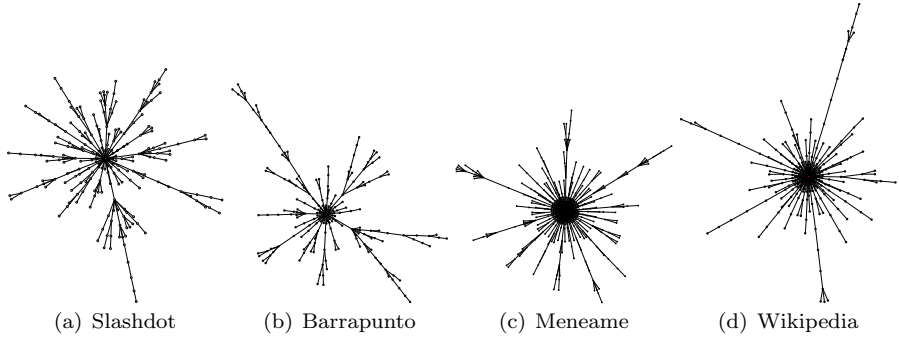

\begin{center}
 \subfigure[Slashdot]{\includegraphics[width=.24\columnwidth]{Fig14.eps}}
 \subfigure[Barrapunto]{\includegraphics[width=.24\columnwidth]{Fig15.eps}}
 \subfigure[Meneame]{\includegraphics[width=.24\columnwidth]{Fig16.eps}}
 \subfigure[Wikipedia]{\includegraphics[width=.24\columnwidth]{Fig17.eps}}
\end{center}
\caption[Example]{Examples of synthetic discussion threads.}
\label{fig:synt}
\end{figure*}

To conclude this section, we show in Figure \ref{fig:synt} the synthetic
counterpart of Figure \ref{fig:real}, where we plot representative threads
with similar sizes selected randomly from each of the four synthetic datasets.
We can see that the generated threads present a strong
resemblance with the real ones.

\subsection{Model validation: evolution of the threads}
\label{sec:evolution}
After having compared the main structural properties of the synthetic
trees with the real ones, we now investigate whether the models
considered in this study are also able to reproduce the growth process
of the threads.  In other words, if we take intermediate snapshots of
the threads during their evolution, how close match the synthetic
trees their archetypes?

\begin{figure}[!t]
\begin{center}
\subfigure[Full Model\label{fig:width_timeFM}]{\includegraphics[angle=-90,width=.49\columnwidth]{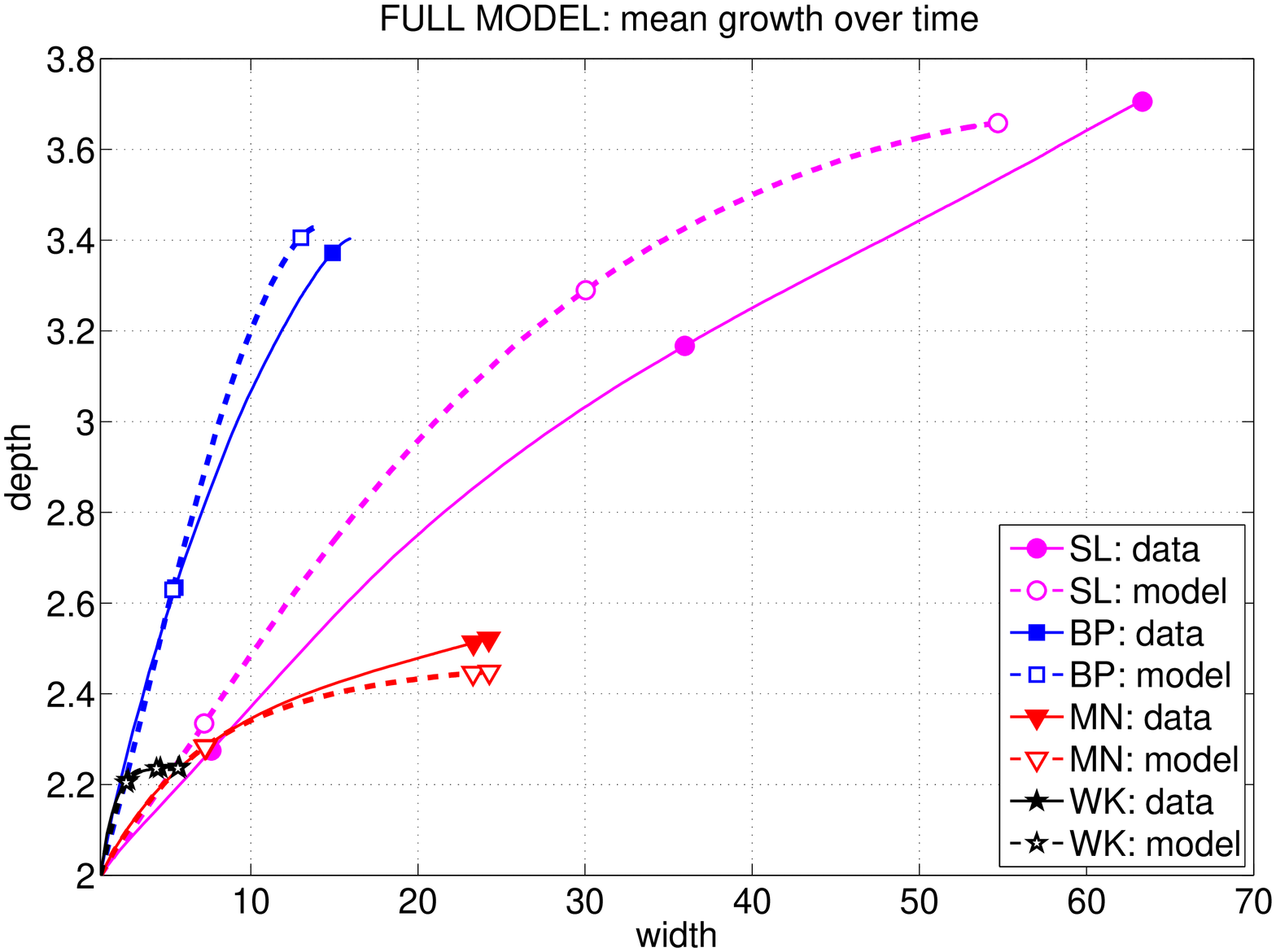}}
\subfigure[NO-$\tau$]{\includegraphics[angle=-90,width=.49\columnwidth]{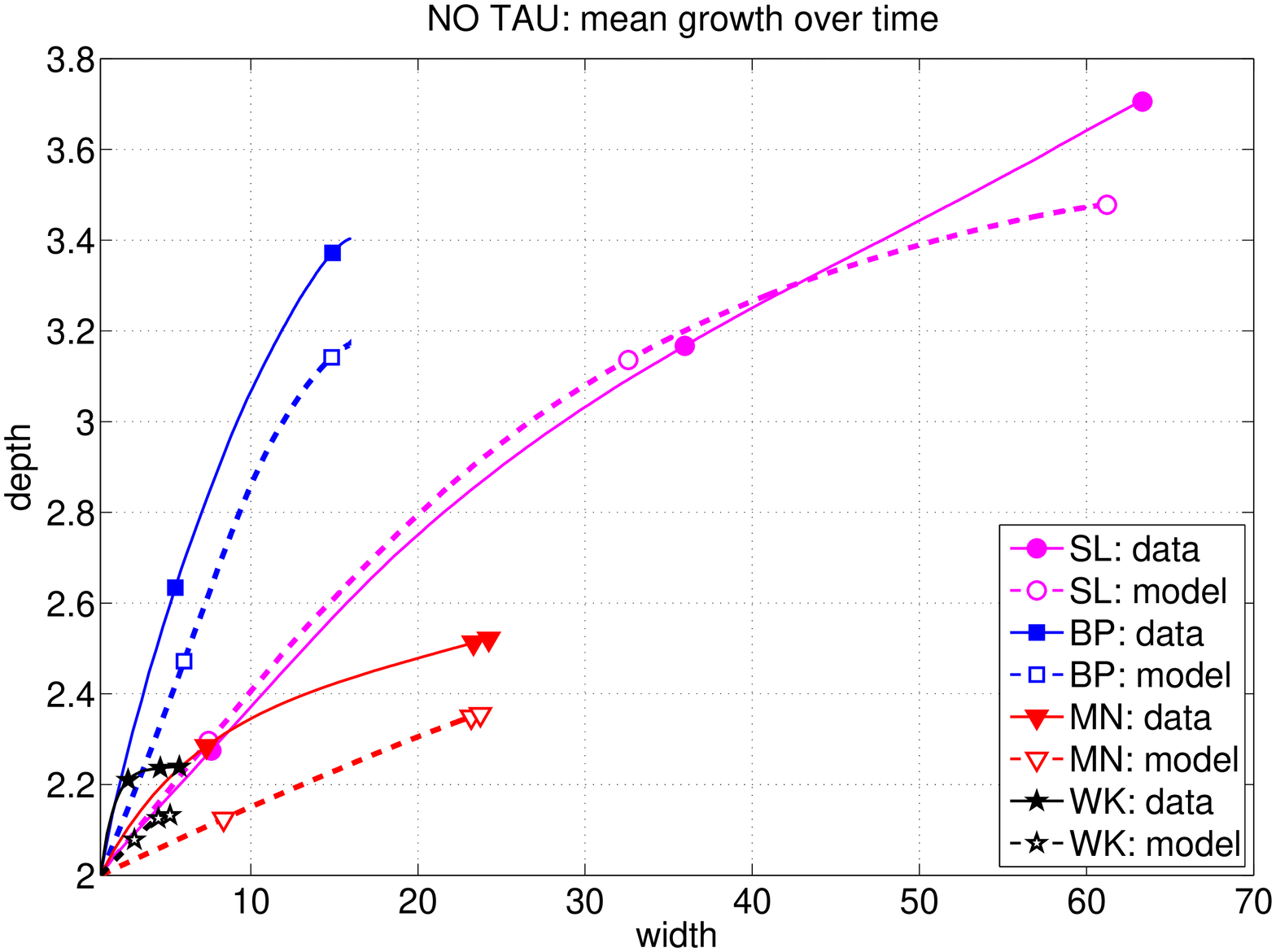}}
\subfigure[NO-bias\label{fig:width_timeNB}]{\includegraphics[angle=-90,width=.49\columnwidth]{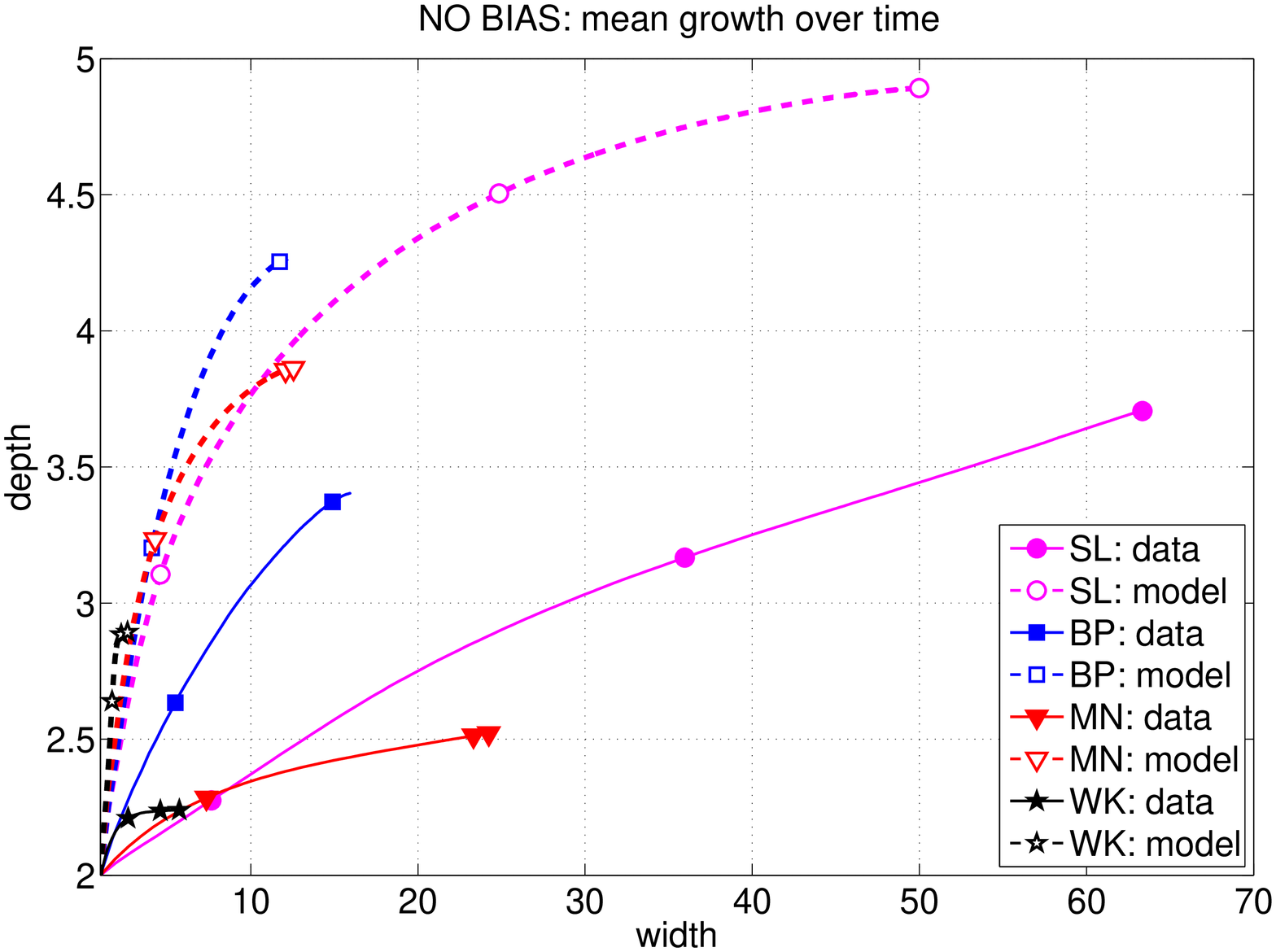}}
\subfigure[NO-$\alpha$]{\includegraphics[angle=-90,width=.49\columnwidth]{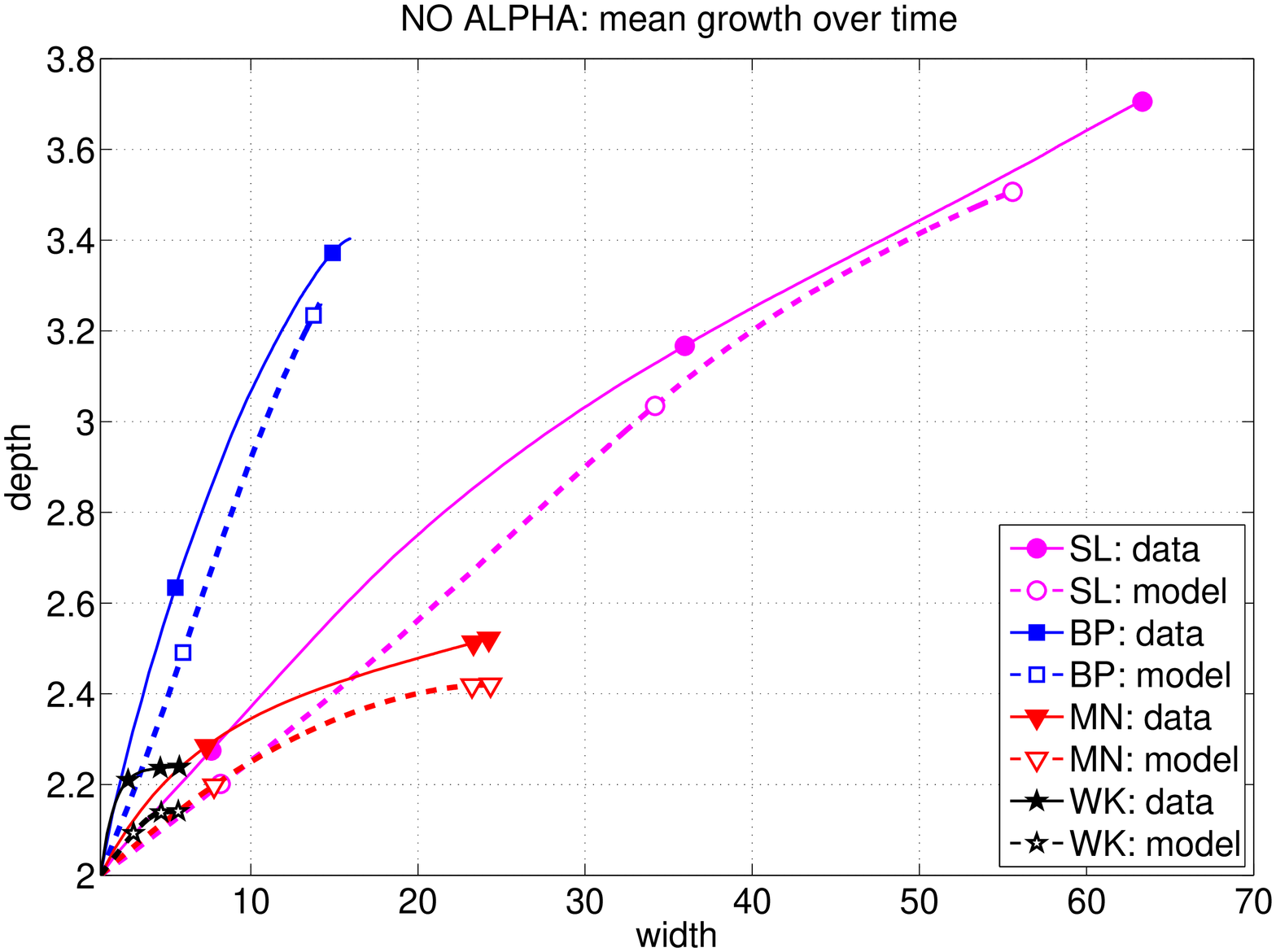}}
\end{center}
\caption{Evolution of the discussions on a
  width-depth plane. Full lines correspond to the discussion obtain
  from our datasets while the dashed lines the growth of synthetic
  discussion generated with one of the four different model types
  analysed in this study. The markers indicate the size of the
  discussions after 10, 100 or 1000 comments. With the exception of
  Slashdot the full model is the one who best approximates the growth
  of the discussions.}
\label{fig:width_time}
\end{figure}

To this end we record two quantities: the \textbf{width} (maximum over the
number of nodes per level) and the \textbf{mean depth} of the trees every time
a new node is added. 
The evolution of the relationship between the averages of these two
quantities is depicted in
Figure~\ref{fig:width_time} 
for all datasets. The marker symbols indicate the size of the
discussions after 10, 100 and 1000 comments, respectively. We compare
the original threads (continuous lines with symbols) with the
different model variants (dashed lines) and observe a good coincidence
between the full model and the data in the evolution of the width of
the discussions (Figure~\ref{fig:width_timeFM}), for three of the four
datasets. Wikipedia shows a nearly perfect coincidence while both,
Meneame and Barrapunto initially follow the growth process of the
synthetic threads but later slightly sub-estimate the mean depth of
the discussions.  However, in the case of Slashdot the full model
generates threads which are deeper but also thinner than the ones
observed in our dataset.

Interestingly, while for all other datasets all three reduced models
reproduce less accurately the growth process of the trees (being
NO-$\alpha$ the second best choice), for Slashdot the NO-$\tau$ model
is initially closer to the evolution of the real threads that the full
model.  How is it possible then that the FM still has a better
log-likelihood than this model?  We can observe that towards the end
of the discussions the full-model produces trees that have nearly the
same average depth as the real instances (albeit in a thiner tree),
while the NO-$\tau$ model produces trees with similar width but
shallower sub-threads. This in the end leads to a better
log-likelihood. And although the actual parameter value of $\tau$ is
very close to one (compare with Table~\ref{tb:params}), it still has an
influence in large discussions where it finally leads to greater
depths, by favouring more recent replies. It seems that in the
specific case of this dataset a two step model would be a better choice
where the effect of $\tau$ would set in only when the discussions are
already of a certain size.

We also observe in coincidence with the reflection on the parameter
$\beta$ at the end of Section~\ref{sec:comparison} that Barrapunto and
Wikipedia have the largest initial growth (largest initial gradient
between width and depth), while Meneame (and even more Slashdot) cause
broader discussions with more comments to the root node. The nearly
identical initial gradient in the NO-bias model, visible in
Figure~\ref{fig:width_timeNB}, confirms this further.

\section{Discussion}
\label{sec:discussion}
We have presented a statistical analysis and comparison of the structure and
evolution of the different discussion threads associated to three popular news
media websites and the discussion pages of the English Wikipedia.  Without
examining the thread content, our analysis already highlights the
heterogeneities between datasets, which could be conditioned mainly by two
factors, namely, the page design, or platform, and the community of users.
Despite these heterogeneities, we have provided evidence that our proposed
mathematical model can capture many of the structural properties and evolution
profiles of the real threads accounting for the particularities of each
dataset.  To the best of our knowledge, this is the first study that analyses
four large-scale datasets in such a level of detail.

The contrast found between Slashdot and Barrapunto (the Spanish version of
Slashdot) is a clear example of how a diverse community of users can
differently shape the activity of a website. Since both websites share the same
platform, the contrasting activity patterns governing their threads should be
mainly attributed to social factors related to the different user communities.
As we show in subsection \ref{sec:comparison}, the proposed framework provides
a rigorous quantitative basis for this effect, which is succinctly represented
in the degree of relevance of novelty and popularity in both spaces.

Another interesting aspect concerns the ability of the proposed framework to
quantify to a certain extent how the thread patterns are influenced by the
interface and/or platform. In Meneame, our analysis identifies the distinction between
two processes, post and comment reply (what we call root-bias), as the most
fundamental feature among novelty and popularity.  Meneame is also the only
website that provides a flat view of the conversations, in contrast to
Slashdot, Barrapunto and Wikipedia, where threads are displayed hierarchically.
We believe that the strongest relevance of the root-bias in Meneame is mainly caused by
the flat interface, which effectively increases the attractiveness of the post,
especially during the beginning of the discussion.  We postulate that this
result is general and could be present in other forums which provide a flat
view of the conversations as well.

The different motivation between Wikipedia discussion pages and typical
conversations on the news media websites manifests in the role of popularity in
the Wikipedia. Whereas popularity seems to be important in the three news media
websites, it is irrelevant in the growth process of Wikipedia discussions. This
conclusion arises naturally from our framework when we compare the likelihoods
of the data between the full model and the model without popularity, since the
differences in the average likelihoods between the two models are not
statistically significant.

The implications of the work presented in this manuscript are diverse.  At a
fundamental level, we have identified a common model that captures the
heterogeneities found in different web-spaces, suggesting that human
conversations are built following a kind of universal law.  This has
implications in the basic understanding of the communication patterns in large
web-spaces that comprise many-to-many interaction.

On an application side, since the parameter estimates of the model allow for a
figurative description of the communication habits of a website, one could
model the structural differences between conversation threads associated with
different communities which share the same platform.  This would lead to
results similar to the one reported for Barrapunto and Slashdot, but for
pre-defined target user communities.  For instance, in terms of political
orientation, how would the parameterisations differ between left/right -wing
oriented forums?  What is the impact of popularity in a model of threads from a
particular political tendency?  The proposed model would be particularly useful
to answer such questions, because of its robustness and its lack of content
interpretation, since semantic analysis may introduce an additional source of
error.

Since novelty and popularity are relevant features in forum design or
maintenance in collaborative applications, it is easy to devise potential
applications in these contexts.  The proposed approach does not require
extensive amounts of data for parameter estimation, which makes it adequate to
track the evolution of the parameterisation over time.  The associated dynamic
patterns could be used to characterize user community evolution and adapt the
design of a forum accordingly.  One could further change the model parameters
and predict how the structural patterns described in sections
\ref{sec:structure} and \ref{sec:evolution} would change in a particular
web-space.


We would like to address several points that deserve further investigation.
First the proposed framework is simple and general. It could be easily applied
to other types of cascades that have similar tree-like structure, such as
chain-letters or forwarded emails \footnote{The datasets considered here and
the source code for parameter estimation and processing the threads are
available on request.}.  We also note that the popularity measure proposed here
(the degree or number of replies) can be replaced by other measures such as
votes or other types of scores.  It would be also interesting to analyse the
impact of using alternative measures of popularity.  

As a consequence of our bottom-up approach, we have focused solely on
structural aspects of the conversational threads and deliberately discarded
other factors such as the precise timing of each node arrival or the cascade
length.  A direction of further research would include the incorporation of the
size of the cascade in the model.  To what extent does the structure of the
discussions depend on their size?  Do short and long discussions exhibit the
same global pattern at different temporal scales, or do large cascades fan out
deeper and narrower  while small cascades follow a more shallow pattern? Our
work suggests that, rather than on the size, the structure is more determined
by the website.  Similar conclusions would have potential implications in a
broader context.

Finally, as suggested in \cite{tempered}, decay of novelty can be due to
competition (limited resources) or habituation.  One would expect competition
to play a more important role within news media which are more sensitive to
fads and habituation to be more typical in Wikipedia.  It would be interesting
to analyse refinements of the novelty term which incorporate these principles.






 
\appendix
\setcounter{section}{0}

\section{: Asymptotics for the degree distribution in FM.}
\label{app:asympt}
Below we provide the heuristic derivation of the power law
distribution in FM.  We use notations as defined in section
\ref{sec:model}. The exact derivation is possible in the same lines as
in \cite[Chapter 8]{Hofstad}. In the full model we have
\[Z_t=2\alpha t+\beta-2\alpha+\tau(1-\tau^t)/(1-\tau).\]

\begin{theorem} For large $t$ and $k$ we have
\[E[d_{k,t}]=\Theta\left(\left(\frac{t}{k}\right)^{1/2}\right).\]
\end{theorem}

{\bf Proof.} For $k>1$ we get
\begin{align}
\nonumber E[d_{k,t+1}]&=E[d_{k,t}]+E[E[d_{k,t+1}-d_{k,t}|{\bf\pi}_{(1:t-1)}]]=E[d_{k,t}]+\frac{\phi(k)}{Z_t}\\
\label{eq:rec}
&=E[d_{k,t}]+\frac{\alpha E[d_{k,t}]+\tau^{t-k+1}}{2\alpha t+\beta-2\alpha+\tau(1-\tau^t)/(1-\tau).}
\end{align}
Next, we construct the sequences $E[\u{d}_{k,t}]$ and $E[\b{d}_{k,t}]$ such that the average degree can be bounded as follows:
\begin{equation}
\label{eq:bounds}
E[\u{d}_{k,t}]\le E[{d}_{k,t}]\le E[\b{d}_{k,t}].\end{equation}
To this end, we define
\begin{equation}\label{eq:1}\u{d}_{k,k}={d}_{k,k}=\b{d}_{k,k}=1,\end{equation}
and for $E[\u{d}_{k,t}]$ and $E[\b{d}_{k,t}]$ we derive the recursive
equations, which constitute, respectively, the lower and the upper bound for
the recursion (\ref{eq:rec}). The lower-bound recursion is constructed
similarly as in \cite{Kumar}:
\begin{align}
E[\u{d}_{k,t+1}]&=E[\u{d}_{k,t}]+\frac{\alpha E[\u{d}_{k,t}]}{2\alpha t+\beta-2\alpha+\tau/(1-\tau)}
\label{eq:ud}
=E[\u{d}_{k,t}]\,\frac{2\alpha t+\beta-\alpha+\tau/(1-\tau)}{2\alpha t+\beta-2\alpha+\tau/(1-\tau)}
\end{align}
For the upper bound recursion note that the right-most expression in (\ref{eq:rec}) is bounded from above by
\begin{align*}
E[{d}_{k,t}]+\frac{(\alpha +\tau^{t-k+1})E[{d}_{k,t}]}{2\alpha t+\beta-2\alpha}
&=E[{d}_{k,t}]\,\frac{2\alpha t+\beta-\alpha}{2\alpha t+\beta-2\alpha}\left(1+\frac{\tau^{t-k+1}}{2\alpha t+\beta-
2\alpha}\right) \\
&\le  E[{d}_{k,t}]\,\frac{2\alpha t+\beta-\alpha}{2\alpha t+\beta-2\alpha}\,e^{\frac{\tau^{t-k+1}}{2\alpha t+\beta-
2\alpha}}.
\end{align*}
Thus, we define $E[\b{d}_{k,t}]$ as
\begin{equation}
\label{eq:bd}
E[\b{d}_{k,t+1}]=E[\b{d}_{k,t}]\,\frac{2\alpha t+\beta-\alpha}{2\alpha t+\beta-2\alpha}\,e^{\frac{\tau^{t-k+1}}{2\alpha t+\beta-
2\alpha}}.
\end{equation}
With $E[\u{d}_{k,t}]$ and $E[\b{d}_{k,t}]$ defined by (\ref{eq:1}), (\ref{eq:ud}) and (\ref{eq:bd}) the inequalities (\ref{eq:bounds}) clearly hold.

Iterating (\ref{eq:ud}) and applying the Stirling's approximation, for large $t$ and $k$ we obtain
\begin{align*}
E[\u{d}_{k,t}]&=\prod_{s=k}^t \frac{2\alpha s+\beta-\alpha+\tau/(1-\tau)}{2\alpha s+\beta-2\alpha+\tau/(1-\tau)}=
\prod_{s=k}^t\frac{s+\frac{\beta-\alpha+\tau/(1-\tau)}{2\alpha}}{s+\frac{\beta-2\alpha+\tau/(1-\tau)}{2\alpha}}\\&=
\frac{\Gamma\left(t+1+\frac{\beta-\alpha+\tau/(1-\tau)}{2\alpha}\right)\Gamma\left(k+\frac{\beta-2\alpha+\tau/(1-\tau)}{2\alpha}\right)}{\Gamma\left(k+\frac{\beta-\alpha+\tau/(1-\tau)}{2\alpha}\right)\Gamma\left(t+1+\frac{\beta-2\alpha+\tau/(1-\tau)}{2\alpha}\right)}\sim \left(\frac{t}{k}\right)^{\frac{1}{2}},\end{align*}
where $a\sim b$ denoted an asymptotic equivalence of $a$ and $b$.
Analogously, for (\ref{eq:bd}), using that 
\[C:=\prod_{s=k}^t e^{\frac{\tau^{t-k+1}}{2\alpha t+\beta-
2\alpha}}\le e^{\frac{\tau(1-\tau^t)}{1-\tau}}\le e^{\frac{\tau}{1-\tau}}<\infty\]
we get that
\[E[\b{d}_{k,t}]\sim C \left(\frac{t}{k}\right)^{\frac{1}{2}},\]
which, together with (\ref{eq:bounds}), proofs the result. \qed

Let us now count the number $\b{N}_{\ge x}$ of values of $k$ satisfying $E[\b{d}_{k,t}]\ge x$, $x>0$. 
\begin{align*}
\b{N}_{\ge x}&= \sum_{k=1}^t{\bf 1}_{\{E[\b{d}_{k,t}]\ge x\}}\sim \sum_{k=1}^t{\bf 1}_{\left\{C \left(\frac{t}{k}\right)^{\frac{1}{2}}\ge x\right\}}=\sum_{k=1}^t{\bf 1}_{\left\{k\le tC^{2}x^{-2}\right\}}=tC^{2}x^{-2}.
\end{align*}
Similarly, for  the number $\u{N}_{\ge x}$ of values of $k$ satisfying $E[\u{d}_{k,t}]\ge x$ we get
\begin{align*}
\u{N}_{\ge x}&= \sum_{k=1}^t{\bf 1}_{\{E[\u{d}_{k,t}]\ge x\}}\sim \sum_{k=1}^t{\bf 1}_{\left\{ \left(\frac{t}{k}\right)^{\frac{1}{2}}\ge x\right\}}=\sum_{k=1}^t{\bf 1}_{\left\{k\le tx^{-2}\right\}}=tx^{-2}.
\end{align*}
Finally, if ${N}_{\ge x}$ of values of $k$ satisfying $E[{d}_{k,t}]\ge x$ then $\u{N}_{\ge x}\le {N}_{\ge x}\le\b{N}_{\ge x}$. Heuristically, ${N}_{\ge x}/t$ gives the asymptotic fraction of nodes with degree at least $x$ at time $t$, thus we obtain the result (\ref{eq:dd}) and 
conclude that the degree distribution follows a power law with exponent $2$ for the cdf, or $3$ for the pdf.
The formal proof is more involved but will lead to the same result because of the concentration of the martingale probability measure around its mean. 

Note that $C$ is bounded by $\exp\left (\frac{\tau}{1-\tau}\right )$, which ranges from one to infinity, in particular for $\tau=0.9$ the value is $e^9\approx 8.1\times 10^3$ and for $\tau=0.99$ it becomes $\approx 9.89\times 10^{42}$. Thus, although $\tau$ does not affect the power law exponent, it can change the fraction of nodes with degree at least $x$ by several orders of magnitude.

%
%
%
%
%

\section{: Parameter estimation and model validation}
We describe here our procedure to obtain parameter estimates for the
different datasets and to validate the model.

For each dataset, we select \emph{with replacement} a subset of $N$ threads
that we use to learn the parameters. We repeat this procedure for $100$
different random subsets of size $N$.  This bootstrap-based approach has
several advantages: first, it reduces the over-fitting when we are dealing with
a small dataset. Note, however, that this is not a concern in our case since
the number of samples (threads) is much larger than the number of parameters
(features)  in all datasets under consideration.  
Second, it reduces the computational cost and makes the parameter estimation 
problem more tractable.  In our case, for Wikipedia, which contains almost one
million of threads, optimisation on the full dataset was too computationally
demanding, but became feasible for a reduced subset of $N=50\cdot 10^3$
threads. As we will see and we already suggested in section \ref{sec:lik},
estimates are already stable for subset sizes of that order.


The results presented in section \ref{sec:results} are based on outcomes of
this estimation procedure. In particular, we show results for $N=50$ and
$N=5\cdot 10^3$ and averaged likelihoods and parameter estimates over the $100$
random realizations.  To validate the model, we analyse the structure and
evolution of the threads generated with respect to the real ones.  We generate
as many threads as the number of threads for each dataset with sizes
pre-determined drawing a pseudo-random number from the empirical distribution
of cascade sizes (see Figure \ref{fig:sizes}). 

%
%

\begin{acknowledgements}
       We wish to thank David Laniado and Riccardo Tasso for providing the
       pre-processed Wikipedia dataset and \url{Meneame.net} for allowing to access an
       anonymised dump of their database.  We also thank Mohammad Gheshlaghi, Wim Wiegerinck and
       Alberto Llera for useful discussions.
\end{acknowledgements}

\bibliographystyle{spmpsci}      
\bibliography{thesis_anonymous}   

\begin{thebibliography}{10}
\providecommand{\url}[1]{{#1}}
\providecommand{\urlprefix}{URL }
\expandafter\ifx\csname urlstyle\endcsname\relax
  \providecommand{\doi}[1]{DOI~\discretionary{}{}{}#1}\else
  \providecommand{\doi}{DOI~\discretionary{}{}{}\begingroup
  \urlstyle{rm}\Url}\fi

\bibitem{adamicyahoo}
Adamic, L.A., Zhang, J., Bakshy, E., Ackerman, M.S.: Knowledge sharing and
  yahoo answers: everyone knows something.
\newblock In: Proceeding of the 17th international conference on World Wide
  Web, WWW '08, pp. 665--674. ACM, New York, NY, USA (2008)

\bibitem{adar05}
Adar, E., Adamic, L.A.: Tracking information epidemics in blogspace.
\newblock In: Proceedings of the 2005 IEEE/WIC/ACM International Conference on
  Web Intelligence, WI '05, pp. 207--214. IEEE Computer Society, Washington,
  DC, USA (2005)

\bibitem{Bakshy}
Bakshy, E., Karrer, B., Adamic, L.A.: Social influence and the diffusion of
  user-created content.
\newblock In: Proceedings of the 10th ACM conference on Electronic commerce, EC
  '09, pp. 325--334. ACM, New York, NY, USA (2009)

\bibitem{Abhijit}
Banerjee, A.V.: A simple model of herd behavior.
\newblock Quarterly Journal of Economics \textbf{107}(3), 797--818 (1992)

\bibitem{barabasi99a}
Barab\'{a}si, A.L., Albert, R.: Emergence of scaling in random networks.
\newblock Science \textbf{286}(5439), 509--512 (1999)

\bibitem{bennaim}
Ben-Naim, E., Krapivsky, P.L.: Stratification in the preferential attachment
  network.
\newblock Journal of Physics A: Mathematical and Theoretical \textbf{42}(47),
  475,001 (2009)

\bibitem{Sushil}
Bikhchandani, S., Hirshleifer, D., Welch, I.: A theory of fads, fashion,
  custom, and cultural change as informational cascades.
\newblock Journal of Political Economy \textbf{100}(5), 992--1026 (1992)

\bibitem{blasio}
Blasio, B.F., Svensson, A., Liljeros, F.: Preferential attachment in sexual
  networks.
\newblock PNAS \textbf{104}(26), 10,762--10,767 (2007)

\bibitem{Bernheim}
Brush, A.B., Wang, X., Turner, T.C., Smith, M.A.: Assessing differential usage
  of {U}senet social accounting meta-data.
\newblock In: Proc. SIGCHI '05, pp. 889--898. ACM, New York, USA (2005)

\bibitem{flickr}
Cha, M., Mislove, A., Gummadi, K.P.: A measurement-driven analysis of
  information propagation in the {F}lickr social network.
\newblock In: WWW'09, pp. 721--730. ACM, New York, USA (2009)

\bibitem{tempered}
D'Souza, R.M., Borgs, C., Chayes, J.T., Berger, N., Kleinberg, R.D.: {Emergence
  of tempered preferential attachment from optimization}.
\newblock Proceedings of the National Academy of Sciences \textbf{104}(15),
  6112--6117 (2007)

\bibitem{polya}
Eggenberger, F., P\'olya, G..: \"uber die statistik verketteter vorg\"ange.
\newblock Mathemathik und Mechanik \textbf{3}, 279--289 (1923)

\bibitem{fisher}
Fisher, D., Smith, M., Welser, H.T.: You are who you talk to: Detecting roles
  in {U}senet newsgroups.
\newblock In: Proc. HICSS '06. IEEE CS, Washington, USA (2006)

\bibitem{goh06}
{Goh}, K.I., {Eom}, Y.H., {Jeong}, H., {Kahng}, B., {Kim}, D.: Structure and
  evolution of online social relationships: Heterogeneity in unrestricted
  discussions.
\newblock Phys. Rev. E \textbf{73}(6), 066,123 (2006)

\bibitem{golub}
Golub, B., Jackson, M.O.: Using selection bias to explain the observed
  structure of internet diffusions.
\newblock PNAS \textbf{107}(24), 10,833--6 (2010)

\bibitem{gomez08}
G\'{o}mez, V., Kaltenbrunner, A., L\'{o}pez, V.: Statistical analysis of the
  social network and discussion threads in slashdot.
\newblock In: Proceedings of the 17th international conference on World Wide
  Web, WWW '08, pp. 645--654. ACM, New York, NY, USA (2008)

\bibitem{gomezHT}
G\'{o}mez, V., Kappen, H.J., Kaltenbrunner, A.: Modeling the structure and
  evolution of discussion cascades.
\newblock In: Proceedings of the 22nd ACM conference on Hypertext and
  hypermedia, HT '11, pp. 181--190. ACM, New York, NY, USA (2011)

\bibitem{GonzalezBailonJIT2010}
Gonzalez-Bailon, S., Kaltenbrunner, A., Banchs, R.E.: The structure of
  political discussion networks: A model for the analysis of e-deliberation.
\newblock Journal of Information Technology \textbf{25}, 230--243 (2010)

\bibitem{blogs}
G\"otz, M., Leskovec, J., McGlohon, M., Faloutsos, C.: Modeling blog dynamics.
\newblock In: ICWSM (2009)

\bibitem{gruhl}
Gruhl, D., Guha, R., Liben-Nowell, D., Tomkins, A.: Information diffusion
  through blogspace.
\newblock In: Proceedings of the 13th International Conference on World Wide
  Web (WWW~'04), pp. 491--501. ACM Press, New York, USA (2004)

\bibitem{Hofstad}
van~der Hofstad, R.: Random Graphs and Complex Networks.
\newblock Lecture notes, available at:.
\newblock
  \urlprefix\url{http://www.win.tue.nl/~rhofstad/Cap_Sel_Connectivity_in_RG.ht%
ml.}

\bibitem{iribarren}
Iribarren, J.L., Moro, E.: Impact of human activity patterns on the dynamics of
  information diffusion.
\newblock Physical Review Letters \textbf{103}(3), 038,702 (2009)

\bibitem{jeong}
Jeong, H., N\'eda, Z., Barab\'asi, A.L.: Measuring preferential attachment in
  evolving networks.
\newblock Europhys. Lett. \textbf{61}(4), 567 (2003)

\bibitem{Joyce2006}
Joyce, E., Kraut, R.E.: Predicting continued participation in newsgroups.
\newblock Journal of Computer-Mediated Communication \textbf{11}, 723--747
  (2006)

\bibitem{kaltenbrunner_LAWEB2007}
Kaltenbrunner, A., G\'omez, V., L\'opez, V.: Description and prediction of
  slashdot activity.
\newblock In: Proceedings of the 5th {Latin American Web Congress (LA-WEB
  2007)}. IEEE Computer Society, Santiago de Chile (2007)

\bibitem{kaltenbrunner2011NRHM}
Kaltenbrunner, A., Gonzalez, G., De~Querol, R., Volkovich, Y.: Comparative
  analysis of articulated and behavioural social networks in a social news
  sharing website.
\newblock New Review of Hypermedia and Multimedia \textbf{7}(3), 243--266
  (2011)

\bibitem{problems}
Kearns, M., Suri, S., Montfort, N.: An experimental study of the coloring
  problem on human subject networks.
\newblock Science \textbf{313}(5788), 824--827 (2006)

\bibitem{Kumar}
Kumar, R., Mahdian, M., McGlohon, M.: Dynamics of conversations.
\newblock In: SIGKDD '10, pp. 553--562. ACM, New York, USA (2010)

\bibitem{twitter}
Kwak, H., Lee, C., Park, H., Moon, S.: What is {T}witter, a social network or a
  news media?
\newblock In: WWW '10, pp. 591--600. ACM, New York, USA (2010)

\bibitem{Lampe05}
Lampe, C., Johnston, E.: Follow the (slash) dot: effects of feedback on new
  members in an online community.
\newblock In: Proc. GROUP '05, pp. 11--20. ACM, New York, USA (2005)

\bibitem{Laniado2011}
Laniado, D., Tasso, R., Volkovich, Y., Kaltenbrunner, A.: {When the Wikipedians
  talk: Network and tree structure of Wikipedia discussion pages.}
\newblock In: ICWSM-11 - 5th International AAAI Conference on Weblogs and
  Social Media. The AAAI Press (2011)

\bibitem{Laniado2010}
Laniado, D., Tasso, R., Volkovich, Y., Kaltenbrunner, A.: {When the Wikipedians
  talk: Network and tree structure of Wikipedia discussion pages.}
\newblock In: ICWSM-11 - 5th International AAAI Conference on Weblogs and
  Social Media. The AAAI Press (2011)

\bibitem{Lerman}
Lerman, K., Hogg, T.: Using a model of social dynamics to predict popularity of
  news.
\newblock In: Proceedings of the 19th international conference on World wide
  web, WWW '10, pp. 621--630. ACM, New York, NY, USA (2010)

\bibitem{viral}
Leskovec, J., Adamic, L.A., Huberman, B.A.: The dynamics of viral marketing.
\newblock In: EC '06, pp. 228--237. ACM, New York, USA (2006)

\bibitem{Leskovec07}
Leskovec, J., McGlohon, M., Faloutsos, C., Glance, N., Hurst, M.: Cascading
  behavior in large blog graphs: Patterns and a model.
\newblock In: SDM '07 (2007)

\bibitem{Nowell08}
Liben-Nowell, D., Kleinberg, J.: {Tracing information flow on a global scale
  using Internet chain-letter data}.
\newblock PNAS \textbf{105}(12), 4633--4638 (2008)

\bibitem{poisson}
Malmgren, R.D., Stouffer, D.B., Motter, A.E., Amaral, L.A.N.: A poissonian
  explanation for heavy tails in e-mail communication.
\newblock Proc. Natl. Acad. Sci. USA \textbf{47}(105), 18,135--18,158 (2008)

\bibitem{Mcglohon2}
Mcglohon, M., Hurst, M.: Community structure and information flow in {Usenet}:
  Improving analysis with a thread ownership model.
\newblock In: ICWSM (2009)

\bibitem{Nonnecke}
Nonnecke, B., Andrews, D., Preece, J.: Non-public and public online community
  participation: Needs, attitudes and behavior \textbf{1}(6), 7--20 (2006)

\bibitem{peruani}
Peruani, F., Tabourier, L.: Directedness of information flow in mobile phone
  communication networks.
\newblock PLoS ONE \textbf{6}(12), e28,860 (2011)

\bibitem{Preece}
Preece, J., Nonnecke, B., Andrews, D.: The top five reasons for lurking:
  Improving community experiences for everyone.
\newblock Computers in Human Behavior \textbf{2}(20), 201--223 (2004)

\bibitem{rangwala}
Rangwala, H., Jamali, S.: Defining a coparticipation network using comments on
  digg.
\newblock IEEE Intelligent Systems \textbf{25}, 36--45 (2010)

\bibitem{rogers}
Rogers, E.M.: Diffusion of innovations, 5th edn.
\newblock Free Press, New York (2003)

\bibitem{Rudas}
Rudas, A., T\'{o}th, B., Valk\'{o}, B.: Random trees and general branching
  processes.
\newblock Random Struct. Algorithms \textbf{31}, 186--202 (2007)

\bibitem{Sack}
Sack, W.: Discourse diagrams: Interface design for very large-scale
  conversations.
\newblock In: Proc. HICSS '00. Volume 3, p. 3034. IEEE CS, Washington, DC, USA
  (2000)

\bibitem{Sadikov}
Sadikov, E., Medina, M., Leskovec, J., Garcia-Molina, H.: Correcting for
  missing data in information cascades.
\newblock In: WSDM '11, pp. 55--64. ACM, New York, NY, USA (2011)

\bibitem{smith02}
Smith, M.: Tools for navigating large social cyberspaces.
\newblock Commun. ACM \textbf{45}(4), 51--55 (2002)

\bibitem{facebook}
Sun, E., Rosenn, I., Marlow, C., Lento, T.M.: Gesundheit! modeling contagion
  through facebook news feed.
\newblock In: ICWSM. The AAAI Press (2009)

\bibitem{szabo}
Szabo, G., Huberman, B.A.: Predicting the popularity of online content.
\newblock Commun. ACM \textbf{53}, 80--88 (2010)

\bibitem{Wang2011}
Wang, D., Wen, Z., Tong, H., Lin, C.Y., Song, C., Barab\'{a}si, A.L.:
  {Information Spreading in Context}.
\newblock In: 20th International World Wide Web Conference (2011)

\bibitem{wattscascades}
{Watts}, D.J.: {A simple model of global cascades on random networks}.
\newblock Proceedings of the National Academy of Science \textbf{99},
  5766--5771 (2002)

\bibitem{Welser}
Welser, H.T., Gleave, E., Fisher, D., Smith, M.: Visualizing the signatures of
  social roles in online discussion groups.
\newblock Journal of Social Structure \textbf{8}(2), 1--32 (2007)

\bibitem{whittaker}
Whittaker, S., Terveen, L., Hill, W., Cherny, L.: The dynamics of mass
  interaction.
\newblock In: Proceedings of the 1998 ACM conference on Computer supported
  cooperative work, CSCW '98, pp. 257--264. ACM, New York, NY, USA (1998)

\bibitem{lik}
Wiuf, C., Brameier, M., Hagberg, O., Stumpf, M.P.H.: A likelihood approach to
  analysis of network data.
\newblock PNAS \textbf{103}(20), 7566--7570 (2006)

\bibitem{wu2007novelty}
Wu, F., Huberman, B.: {Novelty and collective attention}.
\newblock PNAS \textbf{104}(45), 17,599--17,601 (2007)

\bibitem{Zhongbao2003a}
Zhongbao, K., Changshui, Z.: Reply networks on a bulletin board system.
\newblock Phys. Rev. E \textbf{67}(3), 036,117 (2003)

\end{thebibliography}


\end{document}